\documentclass[12pt]{article}

\usepackage{arxiv}

\usepackage[utf8]{inputenc} 
\usepackage[T1]{fontenc}    
\usepackage{hyperref}       
\usepackage{url}            
\usepackage{booktabs}       
\usepackage{amsfonts}       
\usepackage{nicefrac}       
\usepackage{microtype}      
\usepackage{lipsum}		
\usepackage{graphicx}
\usepackage{subcaption}
\usepackage[cmyk]{xcolor}
\usepackage{tikz} 
\usetikzlibrary{calc,spy,shapes,backgrounds,quotes,arrows.meta}
\usepackage{pgfplots}
\usepackage{bm}
\usepackage{amsmath}
\usepackage{float}
\usepackage{numprint}
\usepackage[square,numbers]{natbib}

\title{Fluid-Structure Interaction Simulation of a Coriolis Mass Flowmeter using a Lattice Boltzmann Method}
\newcommand{\headtitle}{FSI Simulation of a Coriolis Mass Flowmeter using a Lattice Boltzmann Method}

\author{Marc~Haussmann\\
	Lattice Boltzmann Research Group\\
	Karlsruhe Institute of Technology\\
	Karlsruhe 76131, Germany\\
	\texttt{marc.haussmann@kit.edu} \\
    \AND
	Peter~Reinshaus\\
	ROTA YOKOGAWA GmbH \& Co. KG\\
	Wehr 79664, Germany\\
	\texttt{peter.reinshaus@de.yokogawa.com}\\
    \And
	Stephan~Simonis\\
	Lattice Boltzmann Research Group\\
	Karlsruhe Institute of Technology\\
	Karlsruhe 76131, Germany\\
	\texttt{stephan.simonis@kit.edu}\\
	\AND
	Hermann~Nirschl\\
	Department of Mechanical Engineering\\
	Karlsruhe Institute of Technology\\
	Karlsruhe 76131, Germany\\
	\texttt{hermann.nirschl@kit.edu} \\
    \And
    Mathias J.~Krause\\
	Lattice Boltzmann Research Group\\
	Karlsruhe Institute of Technology\\
	Karlsruhe 76131, Germany\\
	\texttt{mathias.krause@kit.edu} \\
}

\begin{document}
\maketitle

\begin{abstract}
In this paper we use a fluid-structure interaction (FSI) approach to simulate a Coriolis mass flowmeter (CMF).
The fluid dynamics are calculated by the open source framework OpenLB, based on the lattice Boltzmann method (LBM).
For the structural dynamics we employ the open source software Elmer, an implementation of the finite element method (FEM).
A staggered coupling approach between the two software packages is presented.
The finite element mesh is created by the mesh generator Gmsh to ensure a complete open source workflow. 
The Eigenmodes of the CMF, which are calculated by modal analysis are compared with measurement data.
Using the estimated excitation frequency, a fully coupled, partitioned, FSI simulation is applied to simulate the phase shift of the investigated CMF design. 
The calculated phaseshift values are in good agreement to the measurement data and verify the suitability of the model to numerically describe the working principle of a CMF. 
\end{abstract}


\section{Introduction}
The exact measurement of mass flow of fluids is important in many branches of technology, for example chemical, oil and gas industry. 
It is needed to control processes and ensure safety, filling batches, inventory and others. 
The Coriolis mass flow meter~(CMF) is an accurate instrument, which is becoming increasingly important in various applications~\cite{wang2014}. 
It consists of one or multiple measuring tubes that are stimulated to vibrate by an electromagnetic pulse generator. 
The fluid to be investigated is directed through the tubes. 
Due to inertia, the Coriolis force causes a phase shift of the vibration, which is detected by sensors on both ends of the system. 
As the mass flow of the conveyed fluid is proportional to the Coriolis force, it can be determined directly.

CMFs have been widely described by analytical and structural models~\cite{belhadj2000,cheesewright1998,hemp1994,kazahaya2010,kutin2002,wang2011}. 
These models have helped to understand the fundamental principle of CMF devices.
Nevertheless, the influence of the fluid was greatly simplified and the practical operation could not be described completely.
Therefore, fluid-structure interaction (FSI) models were developed to realize the operating principle, which means that the fluid motion is affected by the measuring pipe oscillation and the pipe motion in turn by the hydrodynamic forces. 
In recent years, iterative two-way FSI models, which consist of a separated computational fluid dynamics (CFD) solver and a computational structural mechanics (CSM) solver, were applied to simulate CMF.

Bobovnik et al.~\cite{Bobovnik.2005} used two different solvers to simulate a straight tube. Commercially available finite volume code for three dimensional turbulent fluid flow and finite element code for a shell structure were coupled. Five different tube lengths were studied simulating free tube vibration. The results for phase shift and frequency were similar to an analytical Flügge shell and potential flow model.
In 2008, Mole et al.~\cite{Mole.2008} extended the three dimensional numerical model of Bobovnik et al.~\cite{Bobovnik.2005} to deal with forced vibration. 
The study comprises the investigation of meter sensitivity at different Reynolds numbers. 
A maximum decrease of $0.4\,\%$ was observed for the lowest Reynolds number. 
This deviation is known as the low Reynolds number effect.
The same numerical model was used by Bobovnik et al.~\cite{Bobovnik.2013} to study the influence of the design parameters on the installation effects of a CMF. 
Installation effects are measured as change of meter sensitivities from fully developed to disturbed fluid flow. 
Considering a single straight tube the errors vary with sensor positions and decrease with increasing tube length. 
In contrast, Kumar~\cite{Kumar.2009} claimed that a CMF is not sensitive to flow profiles. 
The FSI model of ANSYS-CFX was used to consider a straight single tube. 
The results were quite similar for the shorter tube lengths in comparison to previous studies~\cite{Bobovnik.2005}. 
In contrast, the longer tubes showed a higher deviation, which was attributed to the different resolution. 
By changing the viscosity, the Reynolds number was varied and the deviation in meter sensitivity could be captured. 
It was found that at low Reynolds numbers the oscillating viscous fluid forces become relatively strong and interact with the oscillating Coriolis force, which changes the measurement results.
To further investigate the effect of the Reynolds numbers, Kumar and Anklin \cite{Kumar.2011} investigated a curved double tube CMF with an FSI simulation. 
The meter deviation at low Reynolds numbers were found in good agreement to measurement data.
The low Reynolds number effect was indicated as correctable, if the viscosity of the examined fluid is known.
Also Rongmo and Jian \cite{Rongmo.2013} used the ANSYS-CFX FSI module to study the low Reynolds number effect in a U-tube CMF.
They assumed that arising deviations may be due to those different damping factors.
Damping influences the natural frequency of the tube and was expected to change the meter sensitivity. 
	
The aforementioned studies employ traditional discretization methods like the finite volume method (FVM) for the fluid solver.
Meanwhile, alternative approaches, such as the lattice Boltzmann method (LBM), have received increasing attention. 
Its highly efficient parallel algorithm~\cite{heuveline2009,heuveline2010} and the applicability to a wide range of flow phenomena, e.g. flows in complex geometry~\cite{henn2013,augusto2018} or turbulent flows~\cite{haussmann2019large,haussmann2019direct} offer a high potential. 

One of the first approaches that couple LBM to a structural solver can be found by Scholz et al.~\cite{scholz2006}. 
They propose an anisotropic $p$-adaptive method for elastodynamic problems and show a higher convergence rate in comparison to a uniform $p$-version. 
Especially, the load transfer between the fluid and structural mesh were discussed.
Geller et al.~\cite{geller2006} used a partitioned approach to address the famous two-dimensional FSI benchmark case proposed by Turek and Hron~\cite{turek2006}.
The proposed coupling approach by Geller et al.~\cite{geller2006} leads to consistent quantitative result. 
A further study to validate an LBM solver coupled to a $p$-FEM solver with the Turek and Hron~\cite{turek2006} benchmark was published by Kollmannsberger et al.~\cite{kollmannsberger2009}.
The staggered coupling was shown to be sufficient for simulating the reference case due to the weaker impact of the additional mass effect at small time steps.
In contrast, Li et al.~\cite{valero2018} claimed that the added mass effect has a major influence on accuracy and stability. They shown that the use of a non-staggered coupling approach based on subiterations reduce the effect of artificially added mass. 
Based on the previously mentioned studies~\cite{kollmannsberger2009,geller2006,scholz2006}, Geller et al.~\cite{geller2011} extended the developed FSI approach to address three dimensional benchmark problems.

In contrast, this paper aims to demonstrate the feasibility of a complete open source FSI workflow to simulate a CMF. 
Therefore, OpenLB~\cite{olb13,olbpaper}, an open source implementation of LBM, is coupled to the open source FEM framework Elmer~\cite{elmer}.
The implemented coupling procedure uses a staggered approach.
A modal analysis of the CMF geometry is executed to extract the excitation frequency.
The obtained excitation frequency is applied in a frequency response test to evaluate the transient structural setting. 
The partitioned FSI approach is used to calculate the phase shift.
Both the Eigenfrequencies and the phase shift values are compared to measurement data.
The evaluation and validation of a complex engineering problem with a partitioned FSI approach using LBM is a novelty. 
As a further highlight the new FSI workflow is built on open source frameworks to ensure additional adaptions in the coupling interface.

The paper is structured as follows, Section~\ref{sec:metho} introduces the applied FSI approach covering the fluid and structural models. 
In Section~\ref{sec:setup} the CMF test case is depicted in detail. 
The related modal analysis and the subsequent phase shift calculation results, using the FSI approach, are presented and compared to the measurement data in Section~\ref{sec:results}. 
Finally, Section~\ref{sec:conclusion} summarizes the findings and draws a conclusion.

\section{Methodology}
Firstly, the governing equations for the fluid domain presented. 
Afterwards the LBM and the moving boundary approach is introduced.
Next, the Navier--Cauchy equation and the applied solution routine for the structural domain are shown. 
Finally, the FSI approach to coupling the structural an the fluid domain is presented, including the coupling conditions and implementation details.

\label{sec:metho}
\subsection{Fluid Domain}
\subsubsection{Navier--Stokes Equations}
The incompressible Navier--Stokes equations consist of the continuity equation 
\begin{equation}
\label{eq:NS-cont}
\frac{\partial u^f_\alpha}{\partial x_\alpha}  = 0,
\end{equation}
and the momentum equation, which reads
\begin{equation}
\label{eq:NS-mom}
\frac{\partial u^f_\alpha}{\partial t} + \frac{\partial u^f_\alpha u^f_\beta}{\partial x_\beta}=
\frac{\partial T_{\alpha\beta}^f}{\partial	x_\beta}
- \frac{1}{\rho^f} \frac{\partial p}{x_\alpha},
\end{equation}
where Greek indices obey the Einstein notation, $u^f_\alpha$ is the fluid velocity, $p$ is the  pressure field, $T_{\alpha\beta}^f$ is the fluid stress tensor and $\rho^f$ is the fluid density.
Assuming a Newtonian fluid, the fluid stress tensor is given by
\begin{equation}
\label{eq:stressSGS}
T_{\alpha\beta}^{f} = \nu^f\left(\frac{\partial u^f_\alpha}{\partial x_\beta} + \frac{\partial u^f_\beta}{\partial x_\alpha}\right),
\end{equation}
where $\nu^f$ is the kinematic viscosity.

\subsubsection{Lattice Boltzmann Method}
The discretization of the kinetic Boltzmann equation on an equidistant Cartesian grid yields a  finite number of particle distribution functions $f_i$ . The resulting lattice is defined by
$d$ dimensions and $q$ lattice velocities $\bm{c}_i$, $i=0,1,...,q-1$.
In the present work the $D3Q19$ velocity set is applied, which is given by
\begin{equation}
\bm{c}_i=
\begin{cases} 
(0,0,0),  & i=0,\\
(\pm 1, 0, 0),\, (0, \pm 1, 0),\, (0, 0, \pm 1), & i=1,2,...,6,\\
(\pm 1, \pm 1, 0),\, (\pm 1,0 ,\pm 1),\, (0, \pm 1, \pm 1), & i=7,8,...,18.\\
\end{cases}
\end{equation}
The choice of the collision operator is justified by the higher computation performance and the lower memory demand in the used LBM implementation. 
The violation of the rotational invariance~\cite{kang2013} in comparison to $D3Q27$ can be neglected in the laminar flow regime.

The lattice Boltzmann equation without external forces is given by 
\begin{equation} \label{eq:filtered LBM}
f_{i} \left( \bm{x}^{\mathrm{LB}} + \bm{c}_{i} , t^{\mathrm{LB}} + 1 \right) =
f_{i} \left( \bm{x}^{\mathrm{LB}} ,  t^{\mathrm{LB}} \right) \,
+ \Omega_{i},
\end{equation}
where $f_{i}$ is the particle distribution function at discrete lattice position $\bm{x}^{\mathrm{LB}}$ and time step $t^{\mathrm{LB}}$. 
The collision operator $\Omega_i$ is implemented by a single-relaxation time model proposed by Bhatnagar, Gross and Krook~\cite{Bhatnagar1954}. 
It can be defined as
\begin{equation} \label{eq:BGK collision operator}
\Omega_i = -\frac{1}{\tau} \left(f_i (t^{\mathrm{LB}},\bm{x}^{\mathrm{LB}})-f_i^{eq}(\rho^{\mathrm{LB}}, \bm{u}^{\mathrm{LB}})\right),
\end{equation}
where $\tau$ is the relaxation time towards the discrete particle distribution function at equilibrium state $f_i^{eq}$, $\rho^{\mathrm{LB}}$ is the lattice density and $\bm{u}^{\mathrm{LB}}$ the velocity field. 
Hence, the collision operator conserve mass and momentum.
The particle distribution function equilibrium $f_i^{eq}$ is described by a low Mach number truncated Maxwell-Boltzmann distribution 
\begin{equation} \label{eq:Equilibrium distribution function}
f_i^{eq}\left(\rho^{\mathrm{LB}}, \bm{u}^{\mathrm{LB}} \right) 
= \rho^{\mathrm{LB}} \omega_i \left[ 1+\frac{c_{i\alpha} u^{\mathrm{LB}}_{\alpha}}{c_s^2}	
+\frac{u^{\mathrm{LB}}_\alpha u^{\mathrm{LB}}_\beta(c_{i\alpha} c_{i\beta} - c_s^2 \delta_{\alpha\beta})}{2c_s^4} \right],
\end{equation}
where $\omega_i$ are the lattice weights obtained by the Gauss-Hermite quadrature~\cite{He1997,shan2006}, $c_s = 1/\sqrt{3}$ is the speed of sound of the lattice and $\delta_{\alpha \beta}$ is the Kronecker delta.

The discrete moments of the particle distribution functions $f_i$ result in macroscopic flow quantities. 
The density $\rho^{\mathrm{LB}}$, the momentum $\rho^{\mathrm{LB}} \bm{u}^{\mathrm{LB}}$ and the momentum flux $\bm{\Pi}$ are respectively obtained by the zeroth, first and second moments, which are given by
\begin{equation} \label{eq:Compute Rho}
\rho^{\mathrm{LB}} = \sum_{i=0}^{q-1} f_i \; ,
\end{equation}
\begin{equation} \label{eq:Compute Momentum}
\rho^{\mathrm{LB}} \bm{u}^{\mathrm{LB}} = \sum_{i=0}^{q-1} \bm{c}_i f_i,
\end{equation}
\begin{equation} \label{eq:Compute momentum flux}
\bm{\Pi}_{\alpha \beta} = \sum_{i=0}^{q-1} c_{i\alpha} c_{i\beta}  f_i \; .
\end{equation}
The relaxation time $\tau$ is coupled with the lattice kinematic viscosity $\nu^{\mathrm{LB}}$   through
\begin{equation} \label{lattice_nu}
\nu^{\mathrm{LB}} = c_s^2 \left( \tau - 0.5 \right).
\end{equation} 
Taking a simplified isothermal equation of state into account, the lattice pressure is related to the lattice density by 
\begin{equation} \label{lattice_p}
p^{\mathrm{LB}}=c_s^2\rho^{\mathrm{LB}}.
\end{equation} 
The lattice Mach number $\mathrm{Ma}^\mathrm{LB}$ is written as
\begin{equation}
\mathrm{Ma}^\mathrm{LB}=\frac{u^{\mathrm{LB}}_{char}}{c_s},
\end{equation}
where $u^\mathrm{LB}_{char}$ is the characteristic lattice velocity.
In the incompressible limit ($\mathrm{Ma}^\mathrm{LB}\to 0$), the incompressible Navier--Stokes equations (see Eqs.~\eqref{eq:NS-cont} and~\eqref{eq:NS-mom}) are recovered.

Finally, the lattice Boltzmann algorithm is parted into two steps: local collision step and subsequent streaming step. The local collision step is represented by the right-hand side of Eq.~\eqref{eq:filtered LBM} and the streaming step is associated with the left-hand side of Eq.~\eqref{eq:filtered LBM}.

\subsubsection{Moving Boundary Methods}
\label{subsec: mbm}
A fluid-solid interface is required for the simulation of FSI. 
The lattice Boltzmann method typically uses three groups of approaches to describe this type of interface, namely the partially saturated methods~\cite{holdych2003,noble1998}, the immersed boundary methods~\cite{feng2004,kang2011,cheng2014} and the moving boundary methods~\cite{filippova1998,yu2003,lallemand2003}. 
In the present study, the latter type of interface description is applied. 

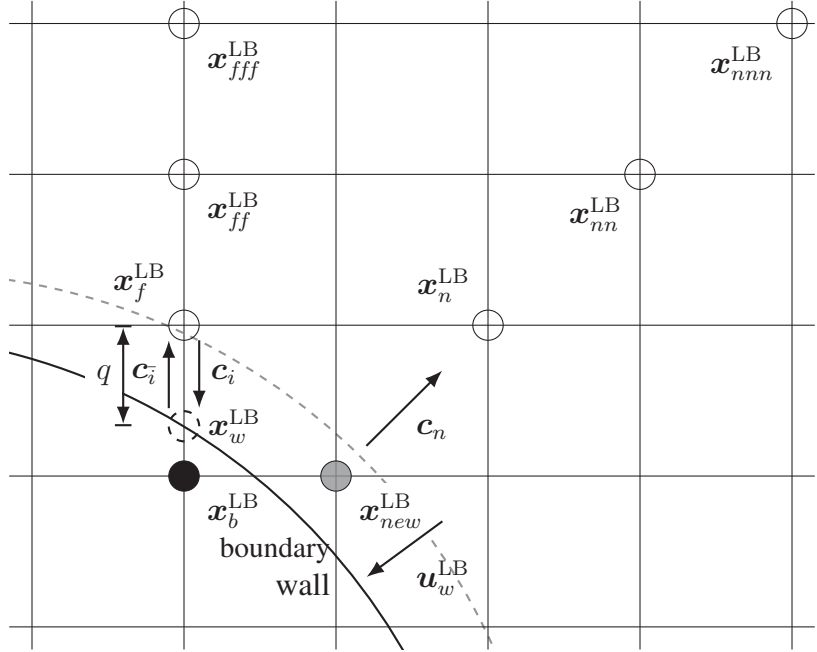
\begin{figure}[htb]
	\centering
	\begin{tikzpicture}[scale=2]
	\definecolor{light-gray}{gray}{0.85}
	\newcommand\radius{0.1}
	\newcommand\shiftC{1.5*\radius}
	\clip (-\shiftC,4+\shiftC) -- (5+\shiftC, 4+\shiftC) -- (5+\shiftC, -\shiftC) -- (-\shiftC, -\shiftC) -- cycle;
	
	\draw[thick] (-1.2,-2.25) circle[radius=4.2]; 
	\begin{scope}[shift={(0.4,0.4)}]
	\draw[thick, gray, dashed] (-1.2,-2.25) circle[radius=4.2]; 
	\end{scope}
	
	\draw (0-\shiftC,0) -- (5+\shiftC,0);
	\draw (0-\shiftC,1) -- (5+\shiftC,1);
	\draw (0-\shiftC,2) -- (5+\shiftC,2);
	\draw (0-\shiftC,3) -- (5+\shiftC,3);
	\draw (0-\shiftC,4) -- (5+\shiftC,4);
	\draw (0,0-\shiftC) -- (0,4+\shiftC);
	\draw (1,0-\shiftC) -- (1,4+\shiftC);
	\draw (2,0-\shiftC) -- (2,4+\shiftC);
	\draw (3,0-\shiftC) -- (3,4+\shiftC);
	\draw (4,0-\shiftC) -- (4,4+\shiftC);
	\draw (5,0-\shiftC) -- (5,4+\shiftC);
	\draw[fill] (1,1) circle[radius=\radius] node [label={[label distance=0.3]-30:$ \boldsymbol{x}_b^\mathrm{LB} $}]{};
	\draw[] (1,2) circle[radius=\radius] node [label={[label distance=0.3]120:$ \boldsymbol{x}_f^\mathrm{LB} $}]{};
	\draw[] (1,3) circle[radius=\radius] node [label={[label distance=0.3]-30:$ \boldsymbol{x}_{f\!f}^\mathrm{LB} $}]{};
	\draw[] (1,4) circle[radius=\radius] node [label={[label distance=0.3]-30:$ \boldsymbol{x}_{f\!f\!f}^\mathrm{LB} $}]{};
	\node at (2,1) [label={[fill=white, label distance=0.3]-30:$ \boldsymbol{x}_{new}^\mathrm{LB} $}]{};
	\draw[fill=gray, fill opacity=0.8] (2,1) circle[radius=\radius];
	\draw[] (3,2) circle[radius=\radius] node [label={[label distance=0.3]120:$ \boldsymbol{x}_n^\mathrm{LB} $}]{};
	\draw[] (4,3) circle[radius=\radius] node [label={[label distance=0.3]-120:$ \boldsymbol{x}_{nn}^\mathrm{LB} $}]{};
	\draw[] (5,4) circle[radius=\radius] node [label={[label distance=0.3]-120:$ \boldsymbol{x}_{nnn}^\mathrm{LB} $}]{};
	
	\draw[align=right] (1.6, 0.4) node{\small boundary\\ wall};	
	
	\draw[thick, dashed] (1.0,1.33) circle [radius=\radius] node [label={[label distance=0.2]0:$ \boldsymbol{x}_w^\mathrm{LB} $}]{};
	
	\begin{scope}[shift={(2.2, 1.2)}]
	\draw[thick,-Latex] (0, 0) -- (0.5,0.5) node[midway, below right] {$ \boldsymbol{c}_{n} $};
	\end{scope}
	\begin{scope}[shift={(1.4, 1.4)}]
	\end{scope}
	\draw[thick, -Latex] (0.9, 1.45) -- (0.9,1.9) node[midway, left] {$ \boldsymbol{c}_{\bar{i}} $};
	\draw[thick, Latex-] (1.1, 1.45) -- (1.1,1.9) node[midway, right] {$ \boldsymbol{c}_i $};
	\begin{scope}[>=Latex]
	\draw[thick, |<->|] (0.6,2) -- (0.6, 1.33) node[midway, left, fill=white] {$ q $};
	\end{scope}
	\draw [thick, -Latex] (2.7,0.7) -- (2.2,0.33) node[midway, below right] {$ \boldsymbol{u}_w^\mathrm{LB} $};
	\end{tikzpicture}
	\caption{Illustration of the used indexing convention.}
	\label{Fig:nodelayout}
\end{figure}

A moving fluid-solid interface inside the fluid domain can be described by the position of its boundary, which changes over time.
The current boundary position indicates fluid and solid nodes. 
If a former solid becomes a fluid node, a refill algorithm is applied to reconstruct the unknown particle distribution functions.
Hence, moving boundary methods are conceptually parted into a velocity boundary formulation and a refill algorithm.
For a better comprehension, the introduced index conventions are displayed in Fig.~\ref{Fig:nodelayout}.
The index $b$ is related to the boundary node positioned inside the solid domain.
In direction $\boldsymbol{c}_i$ the wall is intersected at $\boldsymbol{x}_w^\mathrm{LB}$. The locations $\boldsymbol{x}_f^\mathrm{LB}$, $\boldsymbol{x}_{f\!f}$ and $\boldsymbol{x}_{f\!f\!f}$ denote the corresponding fluid nodes in this direction. 
The distance between $\boldsymbol{x}_w$ and $\boldsymbol{x}_f$ is given by the normalized distance $q$, which is calculated by 
\begin{equation}\label{EQ:q}
q = \frac{|\boldsymbol{x}_f^\mathrm{LB} - \boldsymbol{x}_w^\mathrm{LB}|}{|\boldsymbol{x}_f^\mathrm{LB} - \boldsymbol{x}_b^\mathrm{LB}|}.
\end{equation}
Position $x_{new}^\mathrm{LB}$ indicates nodes, where a refill algorithm is necessary.  
The nodes in discrete normal direction $\boldsymbol{c}_n$ are identified by $\boldsymbol{x}_n^\mathrm{LB}$, $\boldsymbol{x}_{nn}^\mathrm{LB}$ and $\boldsymbol{x}_{nnn}^\mathrm{LB}$. 
This subscript convention is also used for velocity $\boldsymbol{u}^\mathrm{LB}$ and density $\rho^\mathrm{LB}$. 

The present work uses the curved boundary condition proposed by Bouzidi et al.~\cite{bouzidi2001}, which represents an extension of a half-way bounce back boundary scheme.
Thereby a linear interpolation is utilized to take the distance to the boundary into account for increased accuracy.
The unknown populations $f_{\bar{i}}(\boldsymbol{x}_f^\mathrm{LB}, t^\mathrm{LB}+1)$ after the streaming step are calculated by
\begin{equation}
\label{eq:bouzidi:def}
f_{\bar{i}}(\boldsymbol{x}_f^\mathrm{LB}, t^\mathrm{LB}+1) =
\begin{cases}
2q f_i(\boldsymbol{x}_{b}^\mathrm{LB}, t^\mathrm{LB}+1) + (1-2q) f_{i}(\boldsymbol{x}_f^\mathrm{LB},t^\mathrm{LB}+1) -  2 \frac{w_i}{c_s^2} \boldsymbol{c}_i \cdot \boldsymbol{u}_w^\mathrm{LB}(t) & \mathrm{for}\  q < \frac{1}{2}, \\
\frac{1}{2q} f_i(\boldsymbol{x}_{b}, t^\mathrm{LB}+1) + \frac{2q -1}{2q} f_{\bar{i}}(\boldsymbol{x}_{f\!f}^\mathrm{LB},t^\mathrm{LB}+1) - \frac{1}{q} \frac{w_i}{c_s^2} \boldsymbol{c}_i \cdot \boldsymbol{u}_w^\mathrm{LB}(t) & \mathrm{for}\  q \geq \frac{1}{2},
\end{cases}
\end{equation}
where index $\bar{i}$ denotes a quantity in the opposite direction of the one with index $i$.
The half-way bounce back condition is recovered for $ q=1/2 $. 

For the refill algorithm, a second order extrapolation scheme can be found in ~\cite{lallemand2003}
\begin{equation}\label{Eq:extra}
f_i(\boldsymbol{x}_{new}^\mathrm{LB}, t^\mathrm{LB}) = 3 f_i(\boldsymbol{x}_{n}^\mathrm{LB}, t^\mathrm{LB}) - 3 f_i(\boldsymbol{x}_{nn}^\mathrm{LB}, t^\mathrm{LB}) + f_i(\boldsymbol{x}_{nnn}^\mathrm{LB}, t^\mathrm{LB}).
\end{equation} 
Hereby the particle distributions $f_i(\boldsymbol{x}_{new}^\mathrm{LB}, t^\mathrm{LB})$ are extrapolated by the particle distribution functions in discrete normal direction $\boldsymbol{c}_n$.

A further necessary step for FSI is the calculation of the hydrodynamic forces that act on the interface. 
Therefore, a Galilean invariant momentum exchange approach~\cite{wen2014} is used.
The boundary force that acts on a solid node $\boldsymbol{x}_b^\mathrm{LB}$ can be calculated by 
\begin{equation}\label{EQ:ForceWen}
\boldsymbol{F}^\mathrm{LB}(\boldsymbol{x}_b^\mathrm{LB},t^\mathrm{LB}) = \sum_{i\in L} \left[(\boldsymbol{c}_i-\boldsymbol{u}_w^\mathrm{LB}(t))f_i(\boldsymbol{x}_{b}^\mathrm{LB},t+1) - (\boldsymbol{c}_{\bar{i}}-\boldsymbol{u}^\mathrm{LB}_w(t))f_{\bar{i}}(\boldsymbol{x}_f^\mathrm{LB},t^\mathrm{LB}+1)\right],
\end{equation}
where $L$ is the set of fluid-solid links.
This formulation is suitable for the precise description of the boundary force of moving fluid-solid interfaces and avoids the disadvantages of a conventional momentum exchange calculation~\cite{tao2016,wen2014}.

\subsection{Structural Domain}
\subsubsection{Navier--Cauchy Equation}
The present work uses the Navier--Cauchy equation to describe the structural motion.
Therefore, the structral moition is assumed to be linear elastic.
The equation of motion for a linear elastic structure in differential form reads
\begin{equation}
\rho^s\frac{\partial^2\Phi_\beta}{\partial t^2}=\frac{\partial T_{\alpha \beta}^s}{\partial x_\alpha}+\rho^s F_\beta^s,
\end{equation}
where $\Phi_\beta$ is the structural displacement, $\rho^s$ is the solid density, $F_\beta^s$ is the body-force acting on the structure.
Thereby, the Cauchy stress tensor $T_{\alpha \beta}^s$ can be written as
\begin{equation}
T_{\alpha \beta}^s=\mu^s \left(\frac{\partial \Phi_\alpha}{\partial x_\beta} + \frac{\partial \Phi_\beta}{\partial x_\alpha}\right) + \lambda^s \frac{\partial \Phi_\gamma}{\partial x_\gamma}\delta_{\alpha \beta},
\end{equation}
where $\mu^s$ and $\lambda^s$ represents the first and second Lam\'{e} constants.
Both Lam\'{e} constants can be defined by Young's modulus $E$ and the Poisson's ratio $\nu^s$ as
\begin{equation}
\mu^s=\frac{E}{2\left((1+\nu^s\right)}
\end{equation}
and
\begin{equation}
\lambda^s=\frac{\nu^s E}{\left(1+\nu^s\right)\left(1-2\nu^s\right)}.
\end{equation}

\subsubsection{Direct Methods}
This linear Navier--Cauchy equation can be solved by a direct method. 
Therefore, the Elmer solver module is used, which provides the LAPACK collection to address band matrices. 
Direct methods are known for their robustness, but their scaling of order $n^3$ leads to a high memory demand.
Nevertheless, the present work uses a direct method solution procedure due the stability advantages. 
Further information can be found, e.g. in the book of Larson and Bengzon~\cite{larson2013}.   

\subsection{Fluid-Structure Interaction}
\subsubsection{Coupling Conditions}
The FSI problem has to fulfill certain coupling conditions on the interface $\mathcal{I}(t)$, based on physical principles~\cite{richter2017}.
\paragraph{Kinematic condition}
The kinematic condition describes the continuity of the velocities on the interface, i.e.
\begin{equation}
\bm{u}^f(\bm{x},t)=\mathbf{u}^s(\bm{x},t)\quad\text{ on } \mathcal{I}(t).
\end{equation}
The use of Lagrangian and Eulerian coordinate systems for the different solvers requires a mapping procedure.  
\paragraph{Dynamic condition}
The dynamic condition ensures that the forces that act on the interface are balanced due to Newton's third law 'Actio est Reactio'. The coupling condition reads
\begin{equation}
\bm{F}^f(\bm{x},t)=-\bm{F}^s(\bm{x},t)\quad\text{ on } \mathcal{I}(t).
\end{equation}
Due to the different coordinate systems, a mapping procedure is also required here.
\paragraph{Geometric condition}
The third coupling condition is the geometric condition. The condition ensures that the domain is continuous at the interface, i.e.
\begin{equation}
\bm{x}^f(t)=\bm{x}^s(t)\quad\text{ on } \mathcal{I}(t).
\end{equation}
Hence, the fluid and solid domains cannot overlap or separate at the interface. 
\subsubsection{Segregated Approaches}
\label{sec:challenges}
There are several mathematical and technical problems involved in the analysis of FSI.
In most cases, two different subsystems are used for the governing equations.
Even for one subsystem the uniqueness of the solution can be shown only locally in time.
One fundamental problem is the different nature of the partial differential equations. Incompressible Navier--Stokes are of parabolic type, but the structural equation is of hyperbolic type.
Therefore, the different coupling conditions are difficult to ensure on the interface.

\begin{figure}[htb]
\centering
\includegraphics[width=0.7\textwidth]{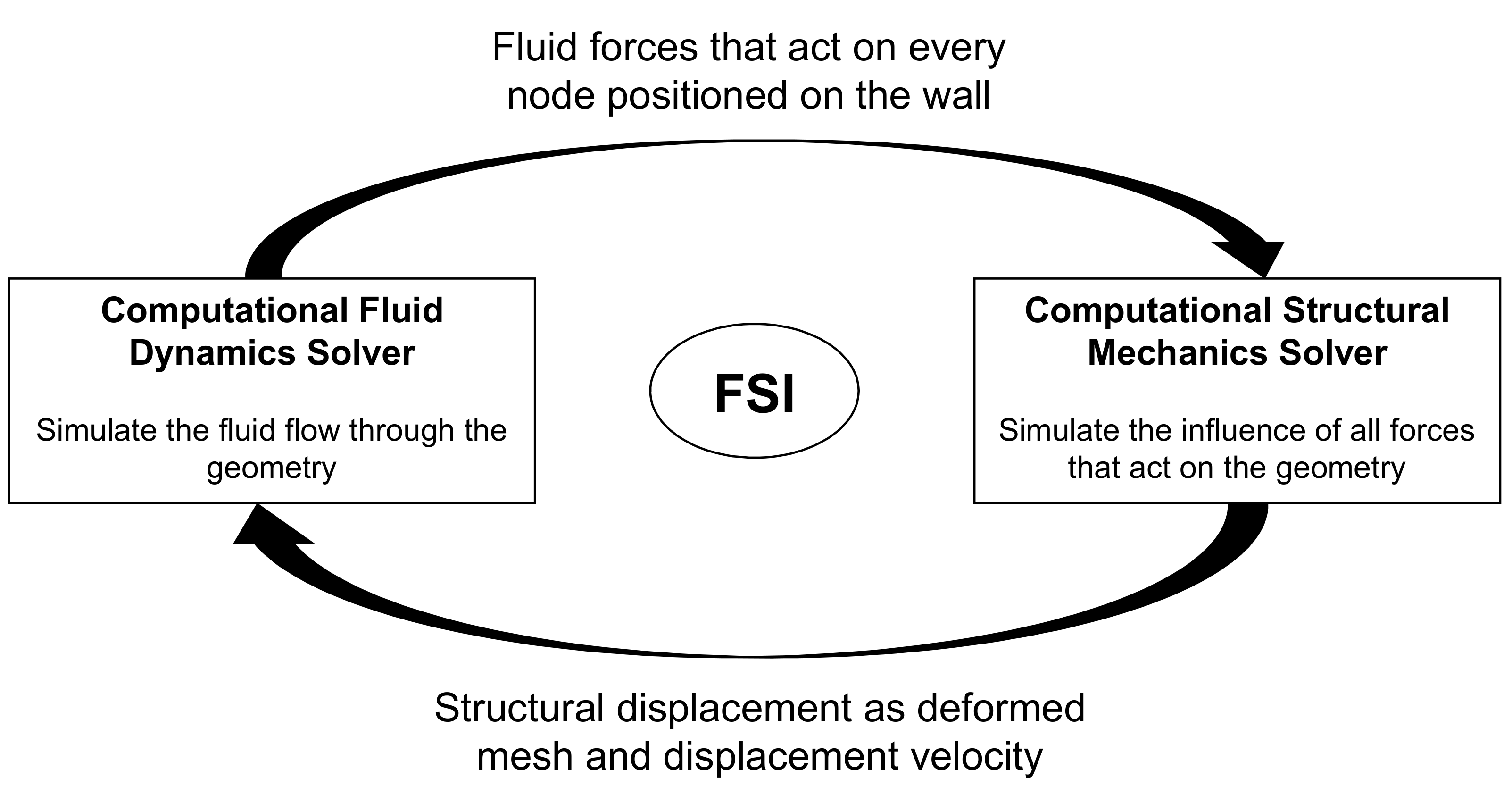}	
\caption{Segregated two-way coupled fluid structure interaction workflow.}
\label{fig:fsi_work_flow}
\end{figure}

Segregated approaches are typically used to address FSI.
The idea is to combine two different solvers, where each solver is specialized to solve either a fluid or a structural problem. 
The coupling is then fulfilled by an outer control instance. 
Due the maturity of each solver, this approach is often a quick possibility to treat complex application problems. 
A common segregated two-way coupled FSI workflow is illustrated in Fig.~\ref{fig:fsi_work_flow}. 
The CFD solver on the left side of the sketch, numerically solves the Navier--Stokes equations on the fluid domain. 
The solution of the fluid field allows to extract the hydrodynamic forces at each grid point in the solid fluid interface that act on the solid. 
This force information is transferred to the CSM solver by an interface. The CSM solver (right side of the sketch) uses the transferred force information as a boundary condition in the structural simulation. The result of the numerical solution of the Navier--Cauchy equation provides the deformed fluid solid interface and the according displacement velocity on each grid point. 
Next, the information is transferred again by the interface operation to the fluid solver. 
The fluid solver in turn, uses the deformed interface and the displacement velocity as a boundary condition. 
This whole process is executed in each coupling period, until a certain time or convergence criterion is fulfilled.

\subsubsection{Implementation}
\label{sec:sim_proced}
\begin{figure}[htb]
\centering
\includegraphics[width=0.8\textwidth]{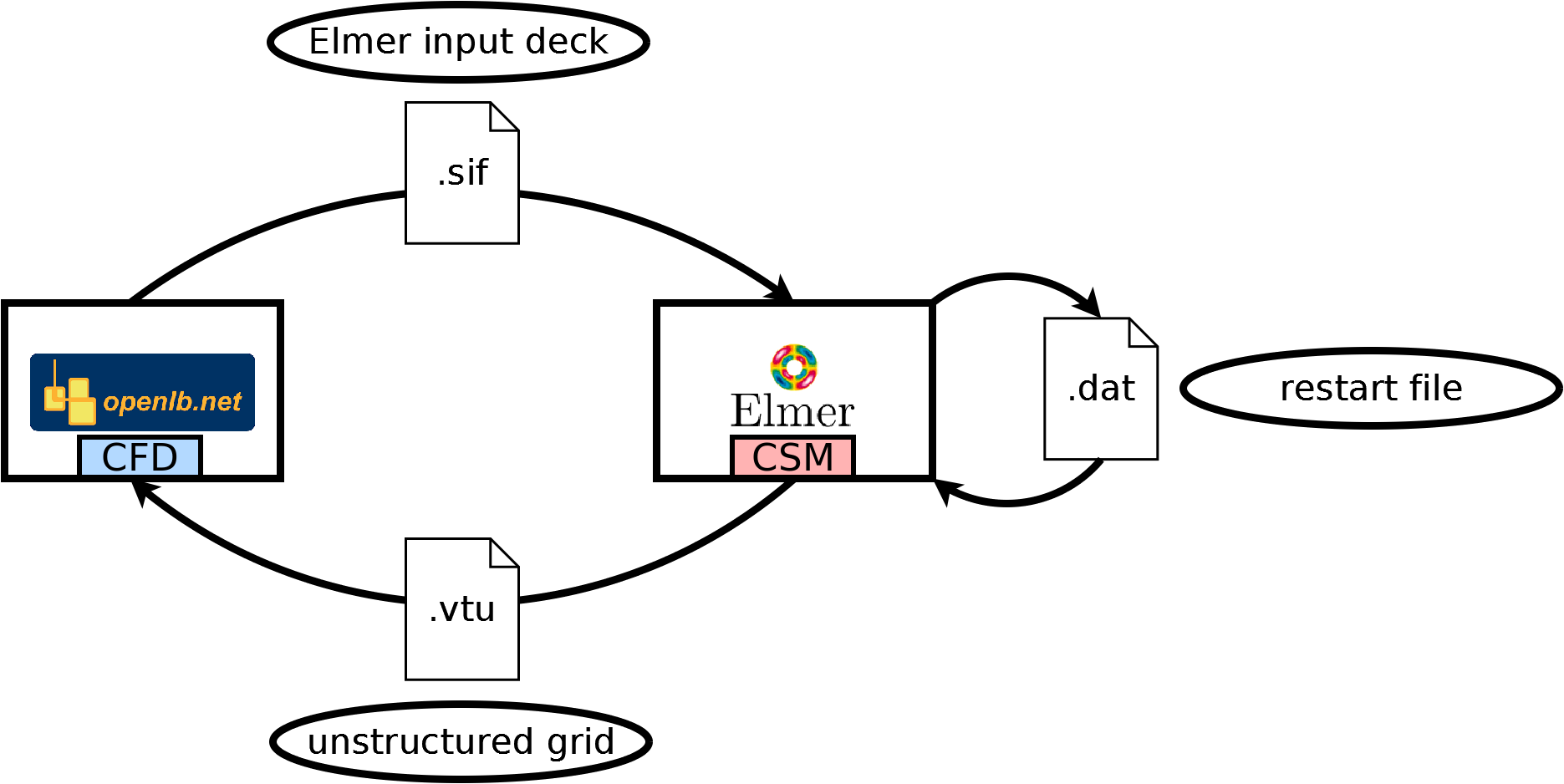}
\caption{Fluid structure interaction workflow using OpenLB and Elmer.}
\label{Elmer_olb_flow}
\end{figure}

The FSI process which uses Elmer and OpenLB is depicted in Fig.~\ref{Elmer_olb_flow}. 
Note that a data based workflow is used to exchange information between the applications.
Currently, the interface allows parallel execution of OpenLB, while Elmer is running in serial mode.
A detailed explanation of each step in the workflow is given in the following procedure, which is executed for every coupling step.
\begin{enumerate}
    \item The OpenLB instance calculates the hydrodynamic forces acting on the boundary for each solid node according to Eq.~\eqref{EQ:ForceWen}.
    \item The hydrodynamic forces are communicated and collected from each worker to the master process.
    \item The master process maps the collected boundary forces to the finite element grid by integrating the force on each finite element mesh point.
    \item The mapped boundary forces are written into an Elmer input deck file (.sif).
    \item Elmer is restarted by the master process using the input deck file (.sif) and a related restart file (.dat).
    \item The Elmer instance is closed after the displacement velocity and the deformed mesh is written to disk as an unstructured mesh file (.vtu) and a new Elmer restart file (.dat) is created.
    \item The master process reads the mesh file (.vtu) and uses the built-in OpenLB voxelizer, which decides whether a point is outside or inside the fluid domain and allows the later distance calculation.
   \item The master process maps the displacement velocity of the FEM grid to the LBM link intersection points $\bm{x}_w$ by a linear interpolation procedure and distributes the information to each worker process.
   \item The OpenLB instance reconstructs the particle distribution functions for the fresh nodes by using the extrapolation refill algorithm (see Eq.~\eqref{Eq:extra}).
   \item The collide and stream algorithm is executed (see Eq.~\eqref{eq:filtered LBM}).
   \item After the streaming step is executed, the unknown particle distribution function are calculated by the curved boundary approach using the mapped displacement velocity (see Eq.~\eqref{eq:bouzidi:def}).
\end{enumerate}

\section{Setup of the Coriolis Mass Flowmeter Test Case}
\label{sec:setup}
\begin{figure}[htb]
\centering
\includegraphics[width=0.5\linewidth]{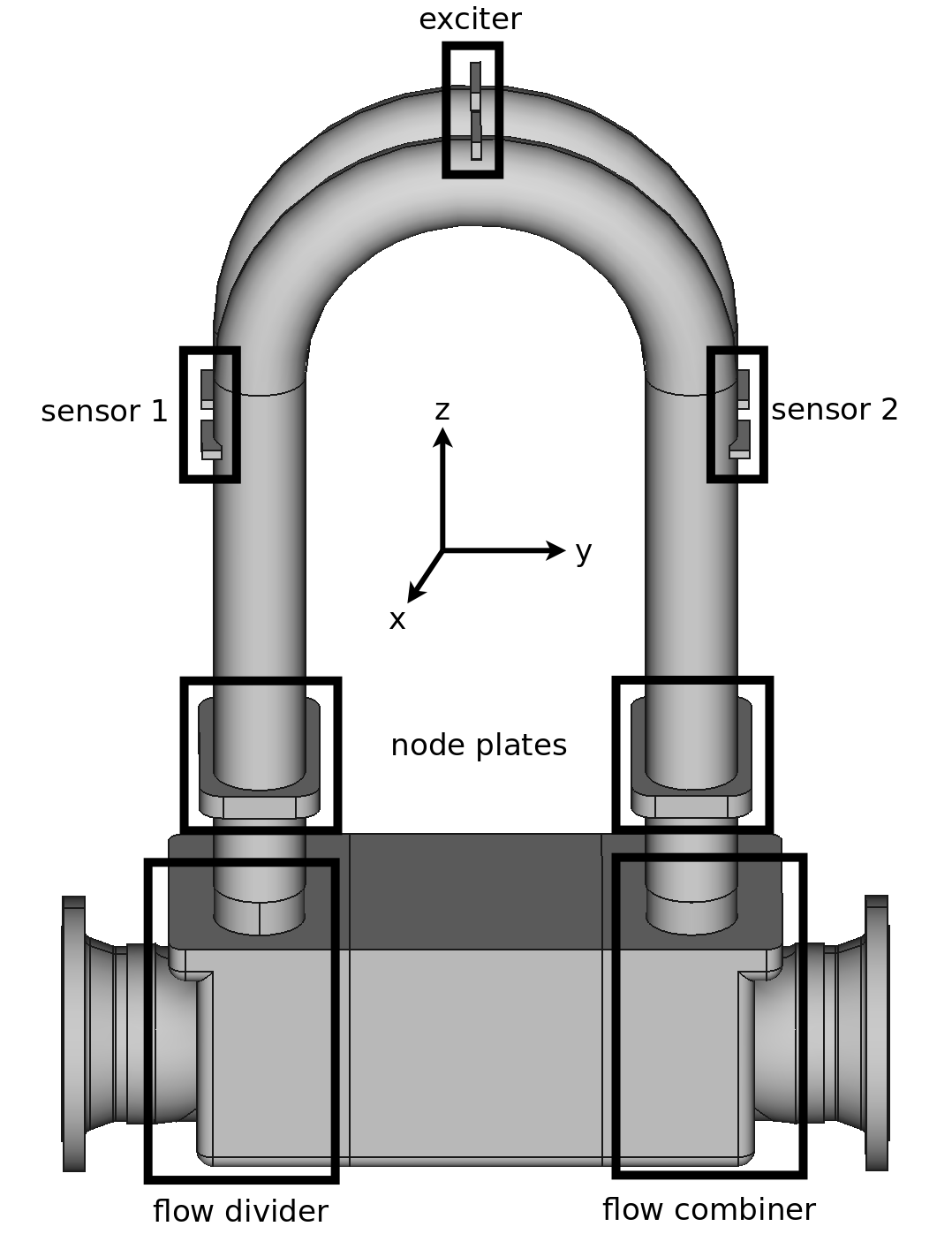}
\caption{Geometry representation and description of the investigated CMF without outer housing.}
\label{geo_desc}
\end{figure}

The investigated CMF geometry is depicted in Figure~\ref{geo_desc}. 
The CMF geometry consists of a flow divider that distributes the incoming mass flow in two U-shaped measuring pipes. 
After the flow passed both measuring pipes, a flow combiner unite the streams. 
The oscillation of the measuring pipe is initialized by an electromagnetic exciter at the top of both measuring pipes.
The resulting oscillation signal is captured at sensor position 1 and 2.  
In addition, two node plates are used to damp the oscillation at the end of the pipes.  
The structural and fluid properties used in the simulations are listed in Table~\ref{tab:fluidstructprop} unless otherwise specified. 
The structural properties correspond to steel. 
The fluid density $\rho^f$ is chosen according to the density of water, but the dynamic viscosity $\eta^f$ is greatly increased to ensure a laminar flow.

\begin{table}[htb]
\centering
\caption{Structural and fluid properties.}
\begin{tabular}{@{}lllllllll@{}}
    \toprule
\multicolumn{2}{l}{Structural properties} &  \multicolumn{2}{l}{Fluid properties} &\\
    \midrule
$\rho^s$ & $\phantom{.}7870\, \mathrm{kg/m^3}$ & $\rho^f$ & $\phantom{0.}998\, \mathrm{kg/m^3}$\\
$E$ & $\phantom{.0}210\, \mathrm{GPa}$ & $\eta^f$ & $0.207\, \mathrm{Pas}$\\
$\nu^s$ & $\phantom{00}0.3$ & &\\
    \bottomrule
  \end{tabular}
\label{tab:fluidstructprop}
\end{table}

\subsection{Boundary Conditions and Initial Conditions}
\subsubsection{Structural Domain}
\label{sec:struc_boundary}
\begin{figure}[htb]
\centering
\includegraphics[width=0.4\textwidth]{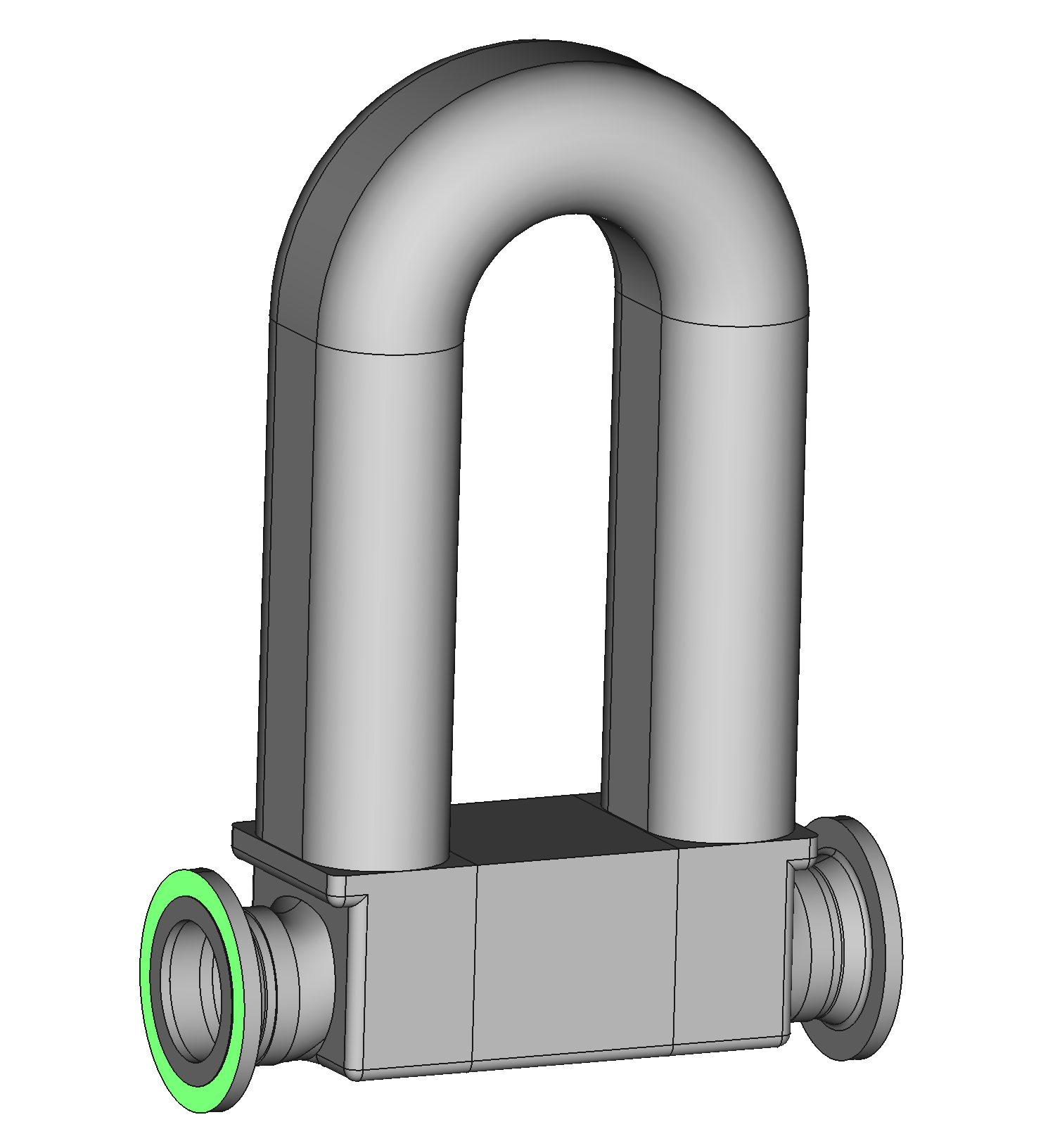}
\caption{Boundary faces at the flanges (green).}
\label{fig:boundary_faces}
\end{figure}

For the structural simulation setting, a zero displacement condition at the flanges is used, i.e.
\begin{equation}\label{eq:flange}
\bm{\Phi}_{flanges}=\bm{0}\,\mathrm{m}.
\end{equation} 
Figure~\ref{fig:boundary_faces} indicates the flange faces in green, where this boundary condition is set.
At the sensor exciter position an excitation load is applied
\begin{equation}
F_x = F_{x,max} \sin (2\pi f_{exc} t) \;\mathrm{for}\; t < \frac{1}{f_{exc}},
\end{equation}
where $F_{x,max}=100\,N$ and $f_{exc}$ is the excitation frequency. 
The force is only acting in the first period to excite the Eigenmode.
\subsubsection{Fluid Domain}
\label{sec:fluid_boundary}
A Dirichlet velocity condition is applied as a boundary condition for the fluid domain at the inlet
\begin{equation}
u^f_{y,inlet}=\left[1-\left(\frac{r}{R} \right)^2 \right]u^f_{y,max},
\end{equation}
where $R$ is the inlet radius and $u_{y,max}$ is the maximum velocity calculated according the used mass flow. This Poiseuille profile assumes a fully developed laminar pipe flow which is justified by an inlet Reynolds number of $\mathrm{Re}_{inlet}<337$.
The pressure on the outlet is set by a Dirichlet condition to 
\begin{equation}
p_{outlet}=0 \,\mathrm{Pa}.
\end{equation}
As FSI is known to be time-consuming it is recommended to start the simulation with a converged flow field. 
Therefore, the fluid geometry is simulated without the structural simulation to initialize the flow field. 
On the measuring pipe walls no-slip boundary conditions are set. 
The flow velocity at the inflow is increased at the inlet for 5 seconds until the desired mass flow is reached. 
This initialization procedure prevents occurring pressure waves due to high gradients and adjusts the non-equilibrium parts of the particle distribution functions.  
\subsubsection{Coupling Conditions}
On each grid point at the interface $\mathcal{I}(t)$ of fluid and solid, the mapped time dependent hydrodynamic force is applied via
\begin{equation}
\bm{F}^s(t) = \bm{F}^f(t) \quad\;\mathrm{on}\; \mathcal{I}(t).
\end{equation}
Thus the dynamic coupling condition is fulfilled.    
The velocity on the fluid structure interface is given by
\begin{equation}
\bm{u}^f(t) = \bm{u}^s(t+\Delta t_c) \quad\;\mathrm{on}\; \mathcal{I}(t+\Delta t_c),
\end{equation}
where $\Delta t_c$ is the coupling period.
The occurring time shift is related to the staggered coupling scheme (see Sec.~\ref{sec:challenges}).
The geometric condition of the interface is also influenced by the coupling period and is written as 
\begin{equation}
\bm{x}^f(t) = \bm{x}^s(t+\Delta t_c)\quad \;\mathrm{on}\; \mathcal{I}(t+\Delta t_c).
\end{equation}

\subsection{Mesh Generation}
\subsubsection{Structural Domain}
\begin{figure}[htb]
\centering
\includegraphics[width=0.5\linewidth]{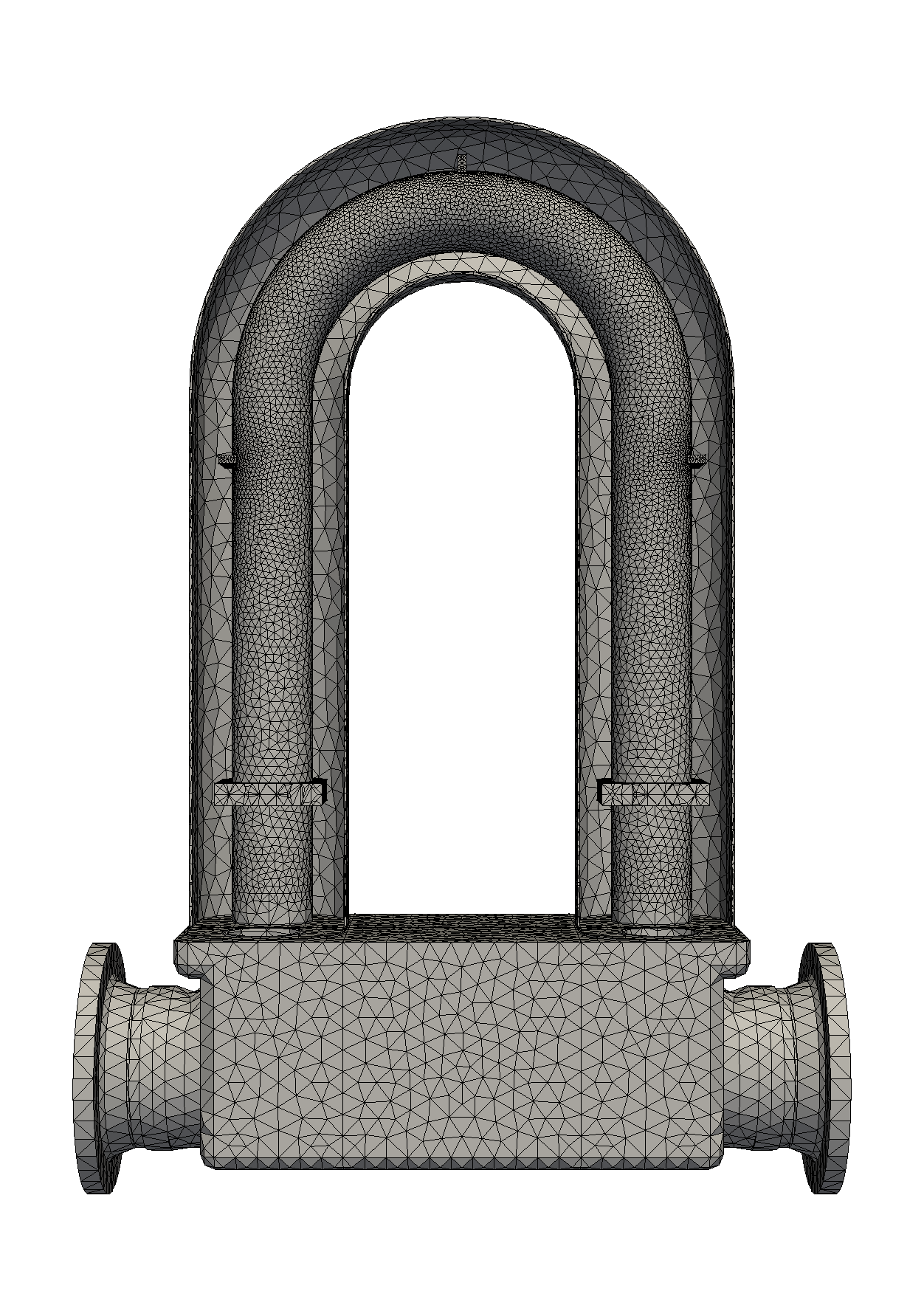}
\caption{Clip representation of the volume mesh.}
\label{Elmer_volume_mesh}
\end{figure}

\begin{figure}[htb]
\begin{subfigure}{0.325\textwidth}
\centering
\includegraphics[width=1.0\textwidth]{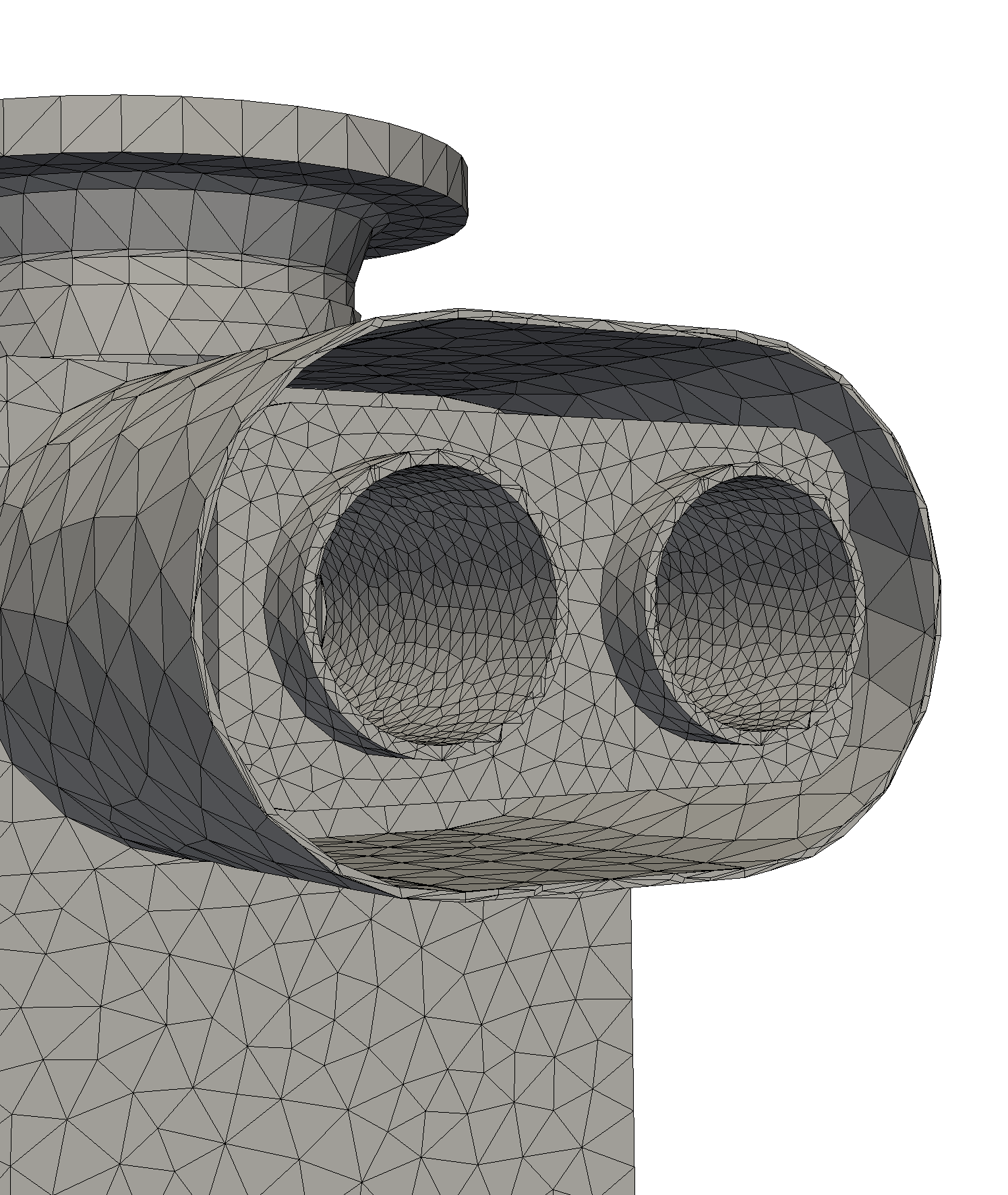}
\subcaption{Node plate -- measuring pipes}
\end{subfigure}
\begin{subfigure}{0.325\textwidth}
\centering
\includegraphics[width=1.0\textwidth]{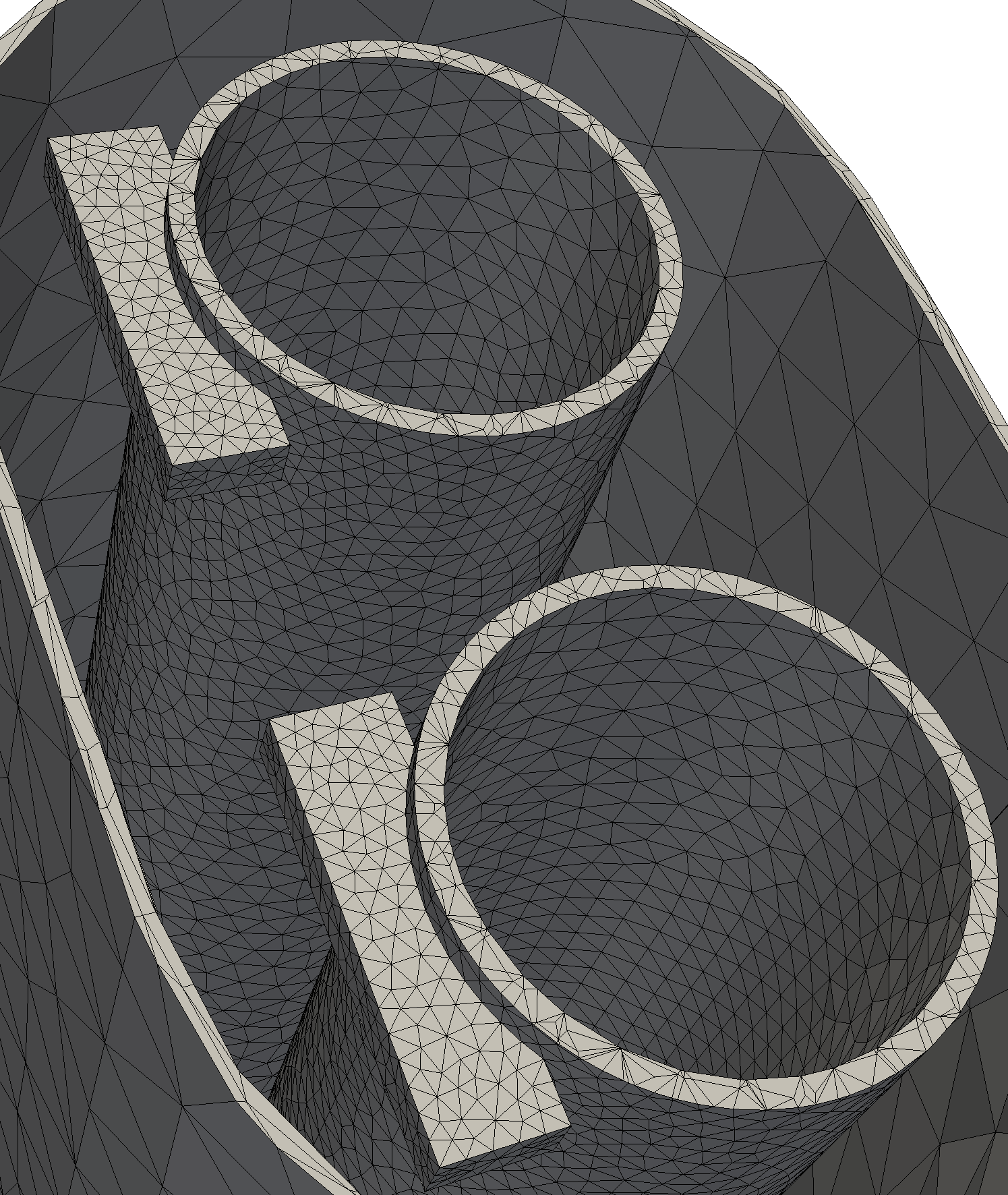}
\subcaption{Sensor -- measuring pipes}
\end{subfigure}
\begin{subfigure}{0.325\textwidth}
\centering
\includegraphics[width=1.0\textwidth]{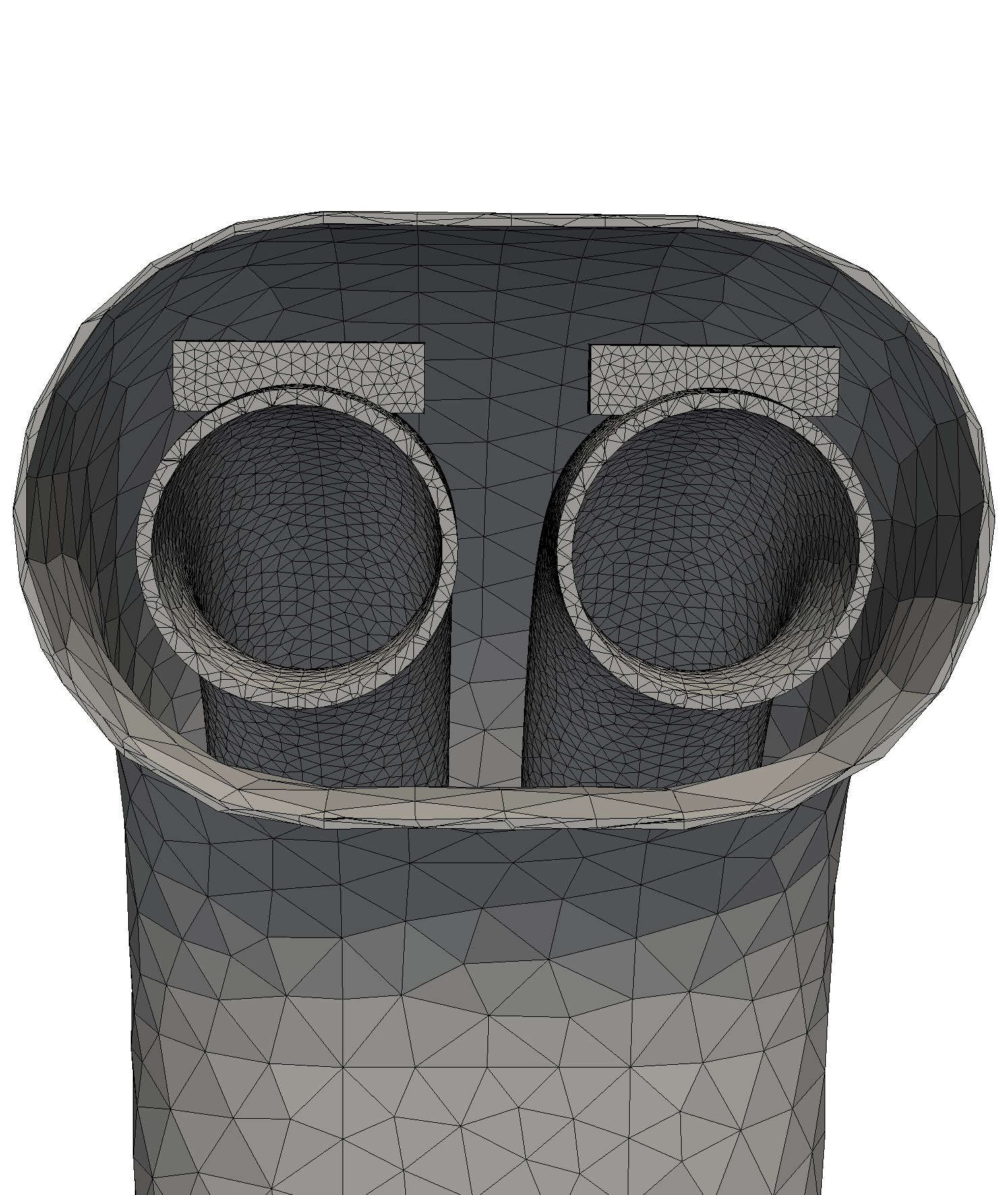}
\subcaption{Exciter -- measuring pipes}
\end{subfigure}
\caption{Connection of the critical mesh regions.}
\label{critical_mesh_regions}
\end{figure}

The computational mesh for the structural simulation is generated by the open source FEM pre-processor Gmsh~\cite{gmsh}. 
The mesh element is chosen according to the geometry pre-processor of OpenLB, because the extraction of a triangulated surface mesh out of tetrahedral volume mesh is straightforward. 
The choice of other mesh element shapes would lead to an additional triangulation step in every coupling period.
In Table~\ref{tab:mesh_ref}, the characteristic length scales of the FEM elements in the according regions are shown.
\begin{table}[htb]
  \centering
  \caption{Averaged characteristic length scales of the finite element mesh regions.}
  \begin{tabular}{@{}lcccccccc@{}}
    \toprule
    Region & $\Delta x^s$ in m\\
    \midrule
Outer housing & 0.035 \\
Body &  0.030 \\
Sensors and exciter & 0.005\\
Measuring pipes & 0.010\\
Node plates & 0.005\\
    \bottomrule
  \end{tabular}
  \label{tab:mesh_ref}
\end{table}
Regions, where simulation results are extracted or high gradients may occur, are refined. 
Therefore, the sensor positions and the measuring pipes require small mesh elements.
The maximal mesh element size is chosen with respect to the largest mesh element size that is used for the housing $\Delta x^s = 0.035\,\mathrm{m}$.
The generated volume mesh in clip representation is depicted in Figure~\ref{Elmer_volume_mesh}.
The mesh contains 52624 nodes and 163164 elements. 
The five predefined regions support different refinement layers. 
This geometry adaptive mesh allows to reduce the amount of mesh points by consistent accuracy of the displacement signal at the sensor positions. 
The quality of the mesh was evaluated with the mesh criteria of Gmsh. 
Furthermore, the connection of critical mesh regions were checked, see Figure~\ref{critical_mesh_regions}.
Particularly in locations where elements are perpendicular to each other, defects may occur.

\subsubsection{Fluid Domain}
\begin{figure}[htb]
\centering
\includegraphics[width=0.5\linewidth]{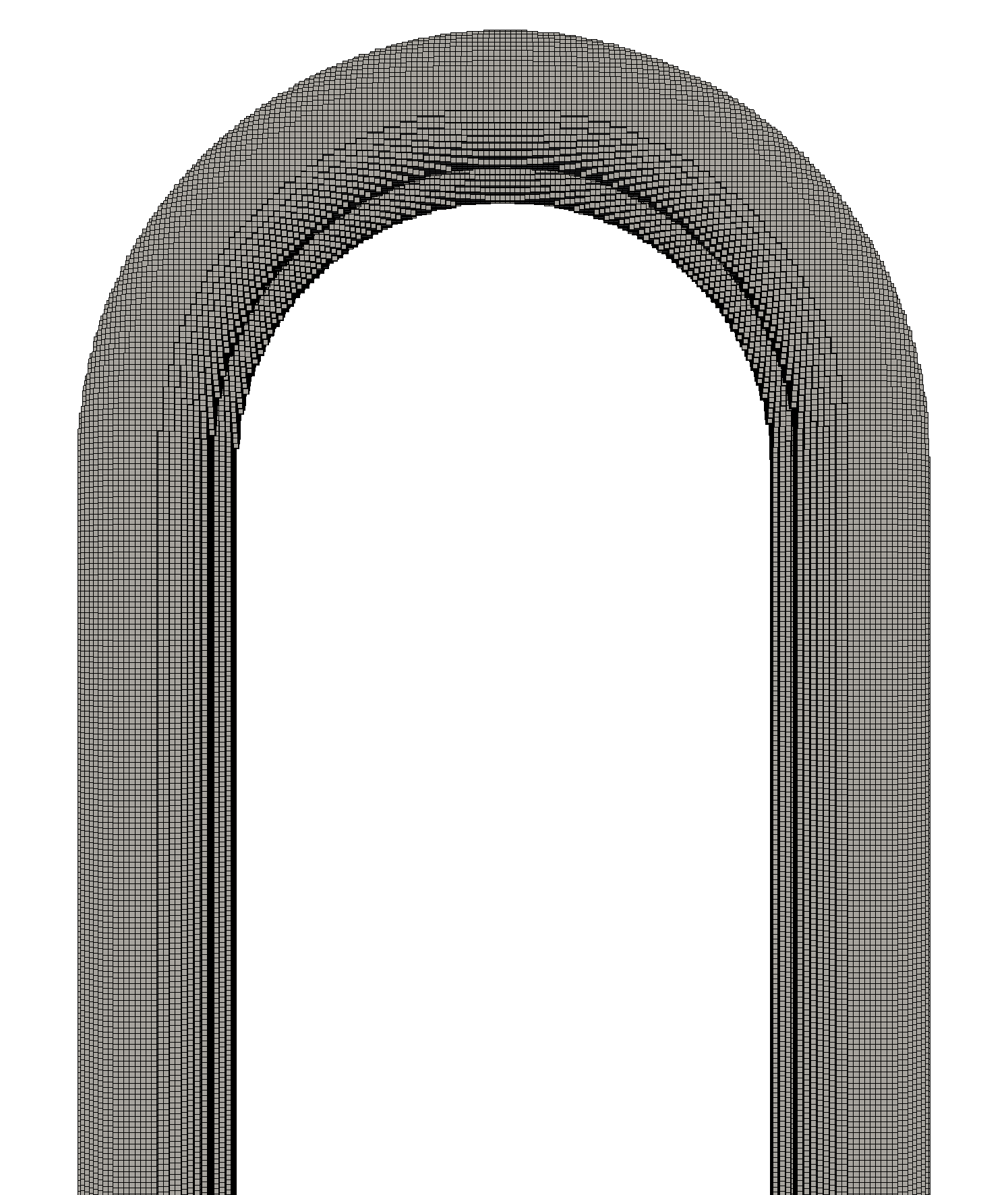}
\caption{Voxelized measuring pipes of the LBM grid.}
\label{fig:lbm_mesh}
\end{figure}

The meshing procedure for LBM is straightforward due to the equidistant uniform Cartesian mesh. 
The used discretization parameters for the two desired mass flows $\numprint{20000}\,\mathrm{\frac{kg}{h}}$ and $\numprint{40000}\,\mathrm{\frac{kg}{h}}$ are shown in Table~\ref{tab:diss_param}. 
\npdecimalsign{.}
\begin{table}[ht]
		\caption{LBM discretization parameters for the both investigated mass flows.}
	\centering
	{\begin{tabular}{@{}ccccccccc@{}}
		\toprule
		Mass flow in $\mathrm{\frac{kg}{h}}$ & \(\Delta x^f\) in m &\(\Delta t^f\) in s &  $\mathrm{Ma^{LB}}$\\
		\midrule 
	    $\numprint{20000}$  & $\numprint{4.056e-3}$ & $\numprint{1.177e-4}$ & $\numprint{8.660e-3}$\\
		$\numprint{40000}$  & $\numprint{4.056e-3}$ & $\numprint{5.885e-5}$ & $\numprint{8.660e-3}$\\
		\bottomrule
	\end{tabular}}
		\label{tab:diss_param}
\end{table}
The resulting grid consists of $1.286$ million grid cells. Figure~\ref{fig:lbm_mesh} shows the voxelized measuring tubes at the beginning of the simulation.
Additional two layers of solid cells cover the measuring pipes to allow the fluid-solid interface motion due to the pipe oscillation.

\section{Results of the Coriolis Mass Flowmeter Test Case}
\label{sec:results}
After the mesh generation is completed, the Eigenfrequencies for the FEM mesh are calculated. 
The detection of the excitation frequency is a preliminary for the later phase shift calculation.
Therefore, a modal analysis is performed with the structural solver Elmer.
\subsection{Modal Analysis} 
The first modal analysis describes the condition for the measuring pipes filled with resting air.
The structural parameters of steel are listed in Table~\ref{tab:fluidstructprop}.
Due to the low density of air compared to steel, the additional mass of air can be neglected. 

\begin{figure}[htb]
\begin{subfigure}{0.24\textwidth}
\centering
\includegraphics[width=1.00\textwidth]{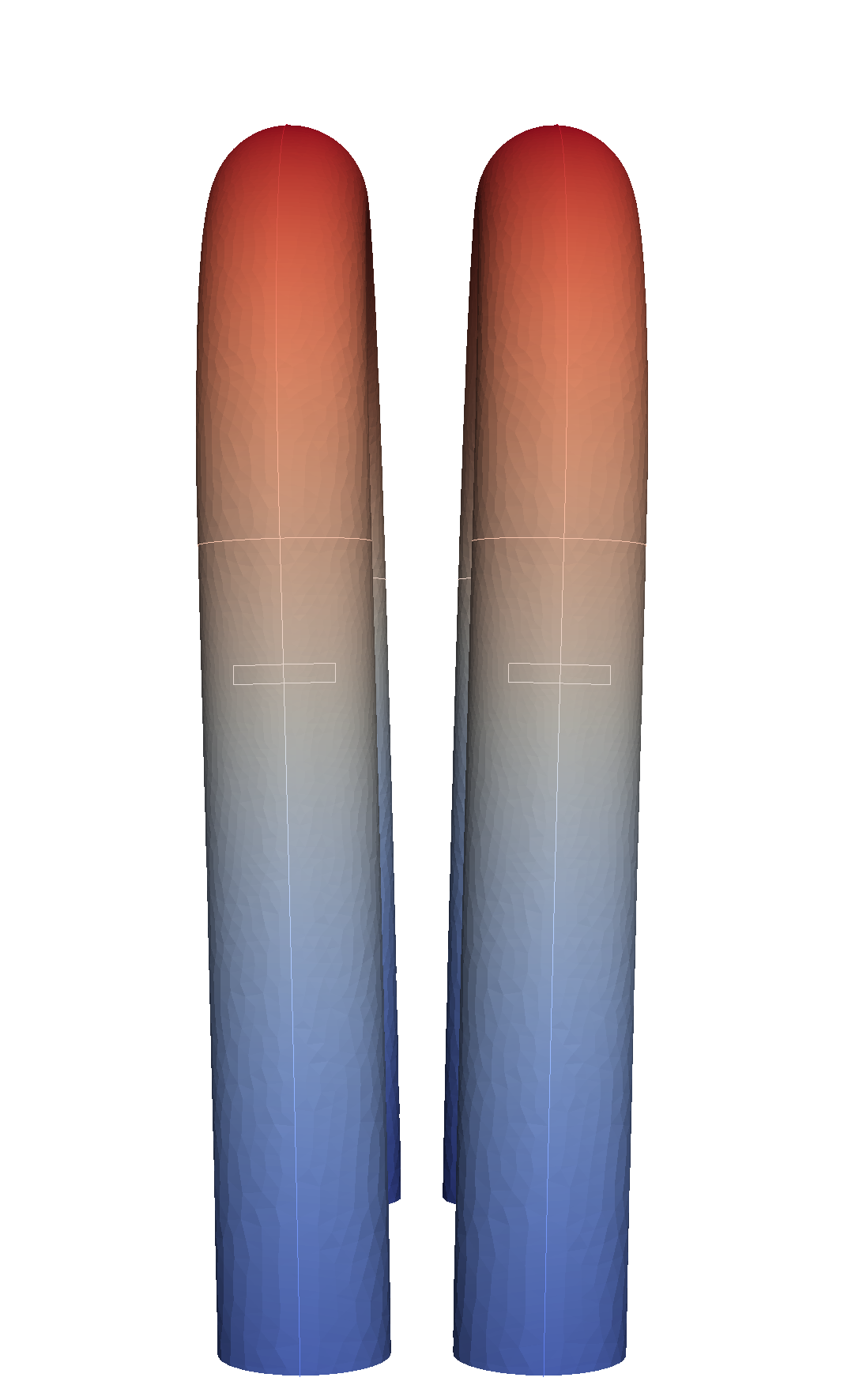}
\subcaption{Mode 2, front view}
\end{subfigure}
\begin{subfigure}{0.24\textwidth}
\centering
\includegraphics[width=1.00\textwidth]{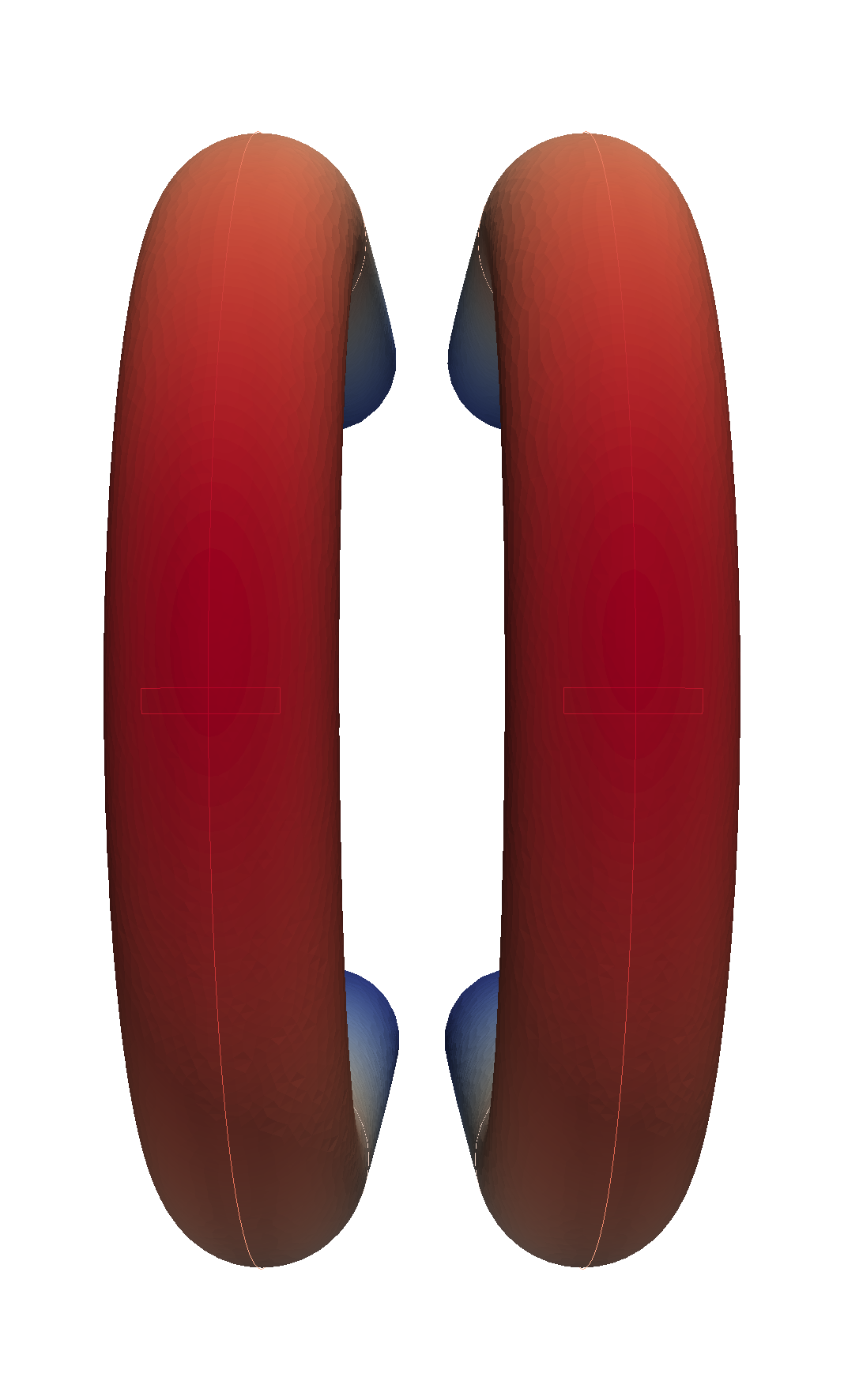}
\subcaption{Mode 2, top view}
\end{subfigure}
\begin{subfigure}{0.24\textwidth}
\centering
\includegraphics[width=1.00\textwidth]{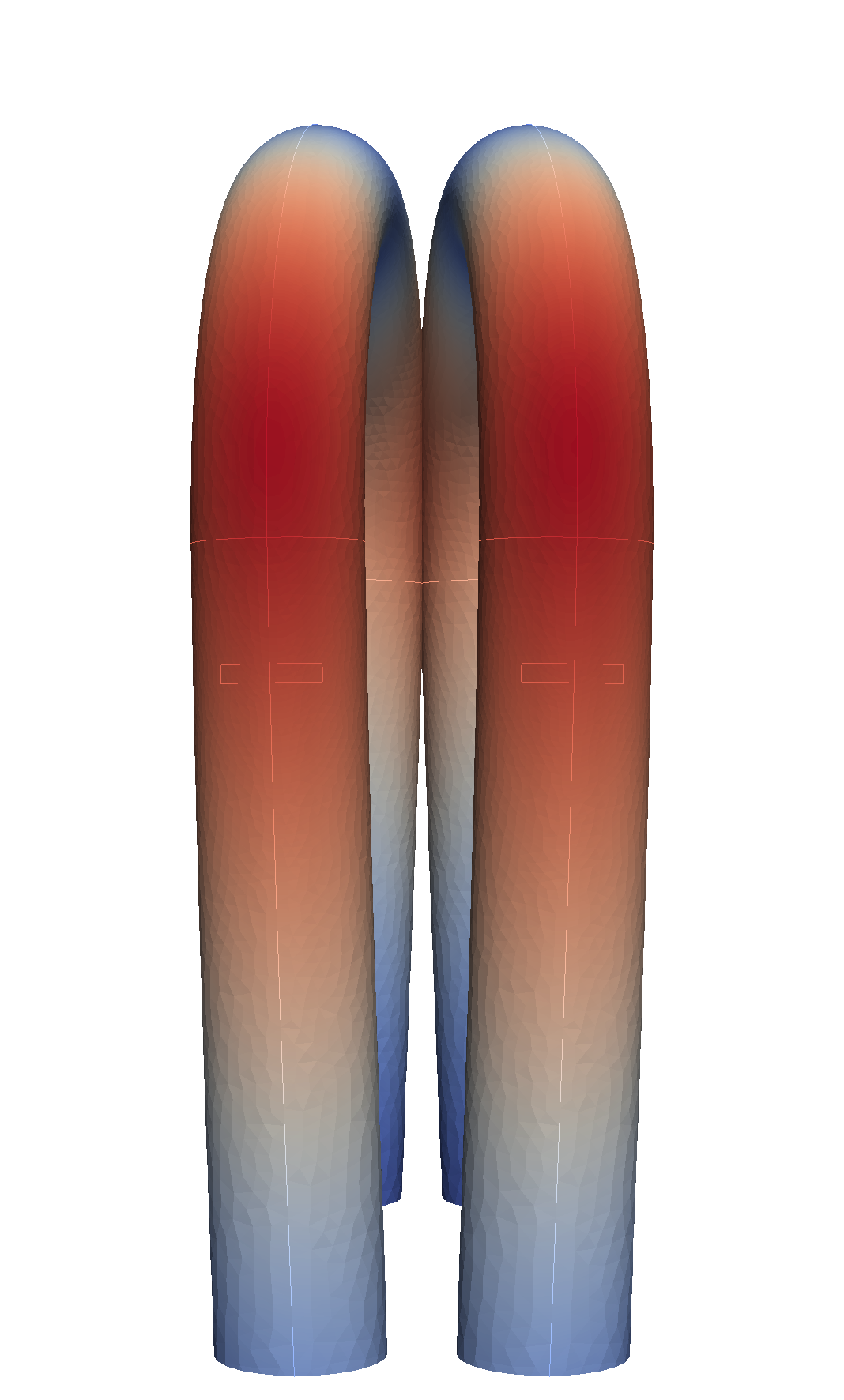}
\subcaption{Mode 8, front view}
\end{subfigure}
\begin{subfigure}{0.24\textwidth}
\centering
\includegraphics[width=1.00\textwidth]{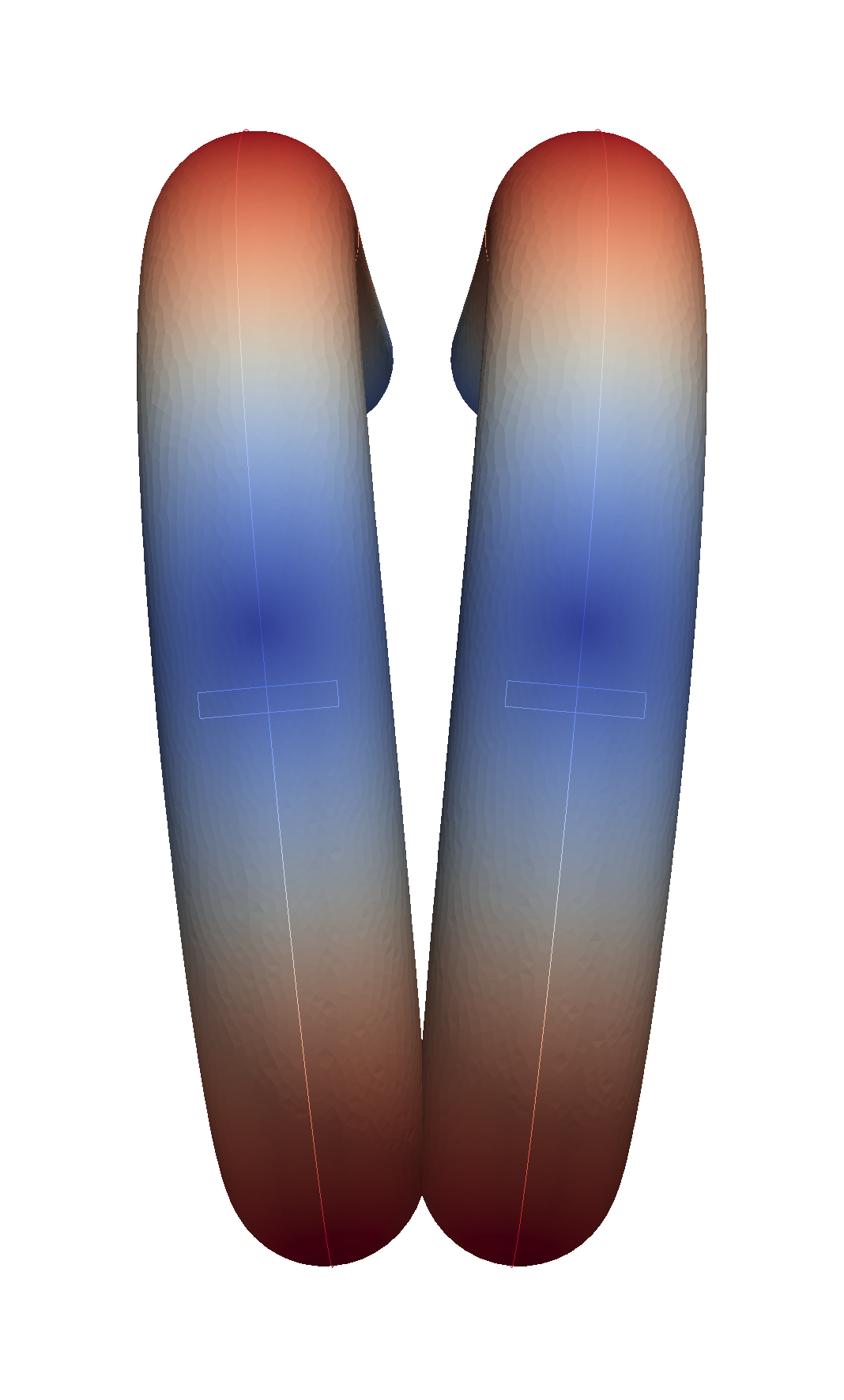}
\subcaption{Mode 8, top view}
\end{subfigure}
\caption{Geometry displacement due to excitation mode (mode 2) and Coriolis twist mode (mode 8).}
\label{fig:top_front_view}
\end{figure}

\begin{figure}[htb]
\centering
\includegraphics[width=0.8\textwidth]{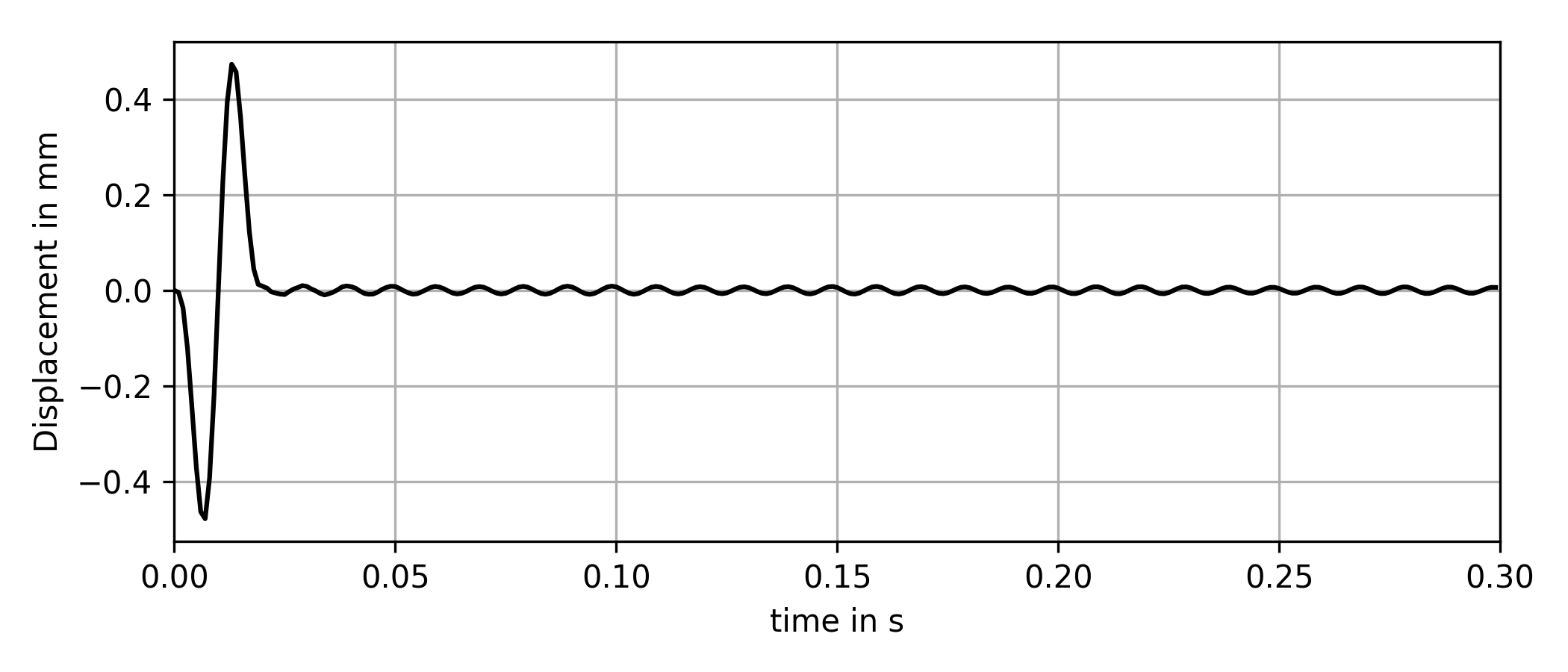}
\caption{Structural response at frequency $f_{exc}=50.00\, \mathrm{Hz}$.}
\label{fig:resp50}
\end{figure}

\begin{figure}[htb]
\centering
\includegraphics[width=0.8\textwidth]{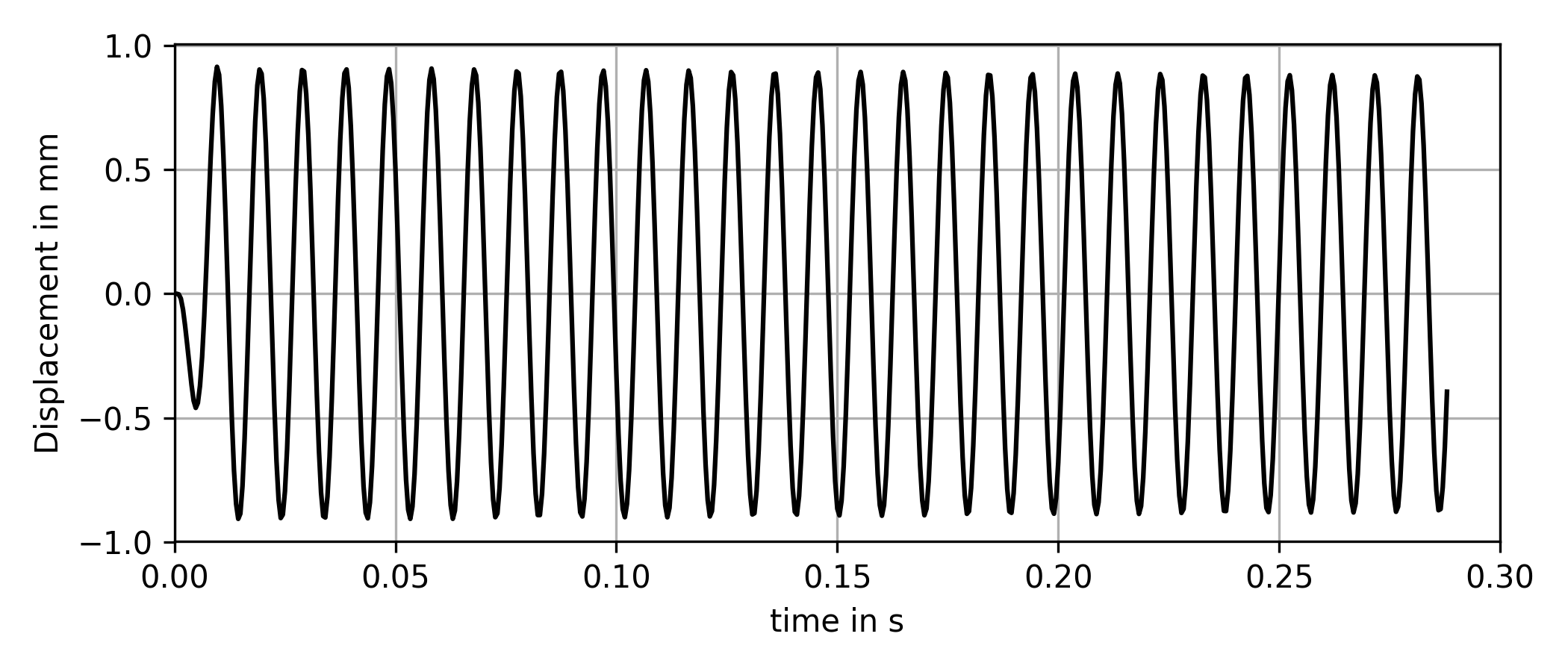}
\caption{Structural response at frequency $f_{exc}=104.28\, \mathrm{Hz}$.}
\label{fig:resp100}
\end{figure}

Using the zero displacement boundary condition (see Eq.~\eqref{eq:flange}), the first ten Eigenfrequencies of the FEM grid are calculated.
The resulting values are shown in Table~\ref{tab:eigen_frequency}.
\begin{table}[htb]
  \centering
  \caption{First ten Eigenfrequencies of the modal analysis and their physical meaning.}
  \begin{tabular}{@{}ccccccccc@{}}
    \toprule
    Mode & $\omega^2$ in $\mathrm{Hz^2}$ & $f$ in $\mathrm{Hz}$ & Physical meaning\\
    \midrule
\phantom{0}1 & $\numprint{2.92e5}$ &\phantom{0}86.02 & \\
\phantom{0}\textbf{2} & $\mathbf{\numprint{4.29e5}}$&	\textbf{104.28} & \textbf{excitation mode}\\
\phantom{0}3 & $\numprint{6.01e5}$ &	123.42&\\
\phantom{0}4 & $\numprint{9.50e5}$ &	155.14&\\
\phantom{0}5 & $\numprint{1.11e6}$ &	167.65&\\
\phantom{0}6 & $\numprint{1.48e6}$ &	193.85&\\
\phantom{0}7 & $\numprint{2.25e6}$ &	238.87&\\
\phantom{0}\textbf{8} & $\mathbf{\numprint{3.03e6}}$ &	\textbf{277.26} & \textbf{Coriolis twist mode}\\
\phantom{0}9 & $\numprint{6.40e6}$ &	402.60&\\
10 & $\numprint{8.05e6}$ &	451.44&\\
    \bottomrule
  \end{tabular}
  \label{tab:eigen_frequency}
\end{table}
A closer look to each Eigenfrequency reveals the physical meaning. 
The searched excitation mode is found at mode number 2 and the Coriolis twist mode corresponds to mode number 8.
The excitation mode is related to a parallel movement of the pipes towards and away from each other. 
On the contrary, the Coriolis twist introduces an additional twist of the pipes.
For a better illustration both modes are displayed in a front and top view in Figure~\ref{fig:top_front_view}.
The next step is the test of the transient structural simulation. Two major aspects are investigated: on the one hand the stability of the transient settings are estimated and on the other hand the resonant behavior are tested. The used structural boundary conditions are described in Section~\ref{sec:struc_boundary}.
In the first case, an excitation frequency different from the Eigenfrequency is selected to $f_{exc}=50\,\mathrm{Hz}$.
In Figure~\ref{fig:resp50}, the structural response over time is plotted.
It can be seen that the amplitude is strongly decreasing after the first period and no resonance is observable. 
This behavior was expected, because the excitation frequency and the Eigenfrequency are mismatched. Nevertheless, the transient simulation is stable over the entire simulation time.
In the second configuration the excitation frequency is chosen with the Eigenfrequency to $f_{exc}=104.28\,\mathrm{Hz}$.  
The displacement signal is depicted in Figure~\ref{fig:resp100}.
The resonance is now clearly visible, which indicates that the results of the modal analysis are reliable and the transient simulation is also stable in the resonant case. 

Additionally a further modal analysis is examined for water conveying tubes, which are used in the FSI case. 
Hereby, the additional mass of water cannot be neglected. 
The water filled tubes are approximated by a fictitious density of the tubes $\rho^s_{fictitious}=\numprint{12319}\, \mathrm{\frac{kg}{m^3}}$, which is calculated by the total mass of the measuring pipes divided by the volume of the structural pipe domain.
The results are summarized and compared to the measurement data in Table~\ref{tab:exc_cor_freq}.
\begin{table}[htb]
  \centering
  \caption{Excitation and Coriolis twist frequency for water and air filled tubes in comparison to measurement data.}
  \begin{tabular}{@{}lcccccccc@{}}
    \toprule
     & Simulation & Measurement & Error in $\%$ \\
    \midrule
$f_{exc,air}$ & 104.28 & 101.00 & \phantom{0}3.24\\
$f_{exc,water}$ & \phantom{0}83.94 & \phantom{0}81.41 & \phantom{0}3.11\\
$f_{Coriolis,air}$ & 277.26 & 249.00 & 11.35\\
$f_{Coriolis,water}$ & 222.92 &	 205.00 & \phantom{0}8.74\\
    \bottomrule
  \end{tabular}
  \label{tab:exc_cor_freq}
  \end{table}
The excitation frequencies for air and water are in good agreement to the measurement data (error $\approx 3\%$).
The errors for the Coriolis frequency seems to be squared due to the higher mode. 

\subsection{Phase Shift Calculation}
\begin{figure}[htb]
\centering
\includegraphics[width=0.8\textwidth]{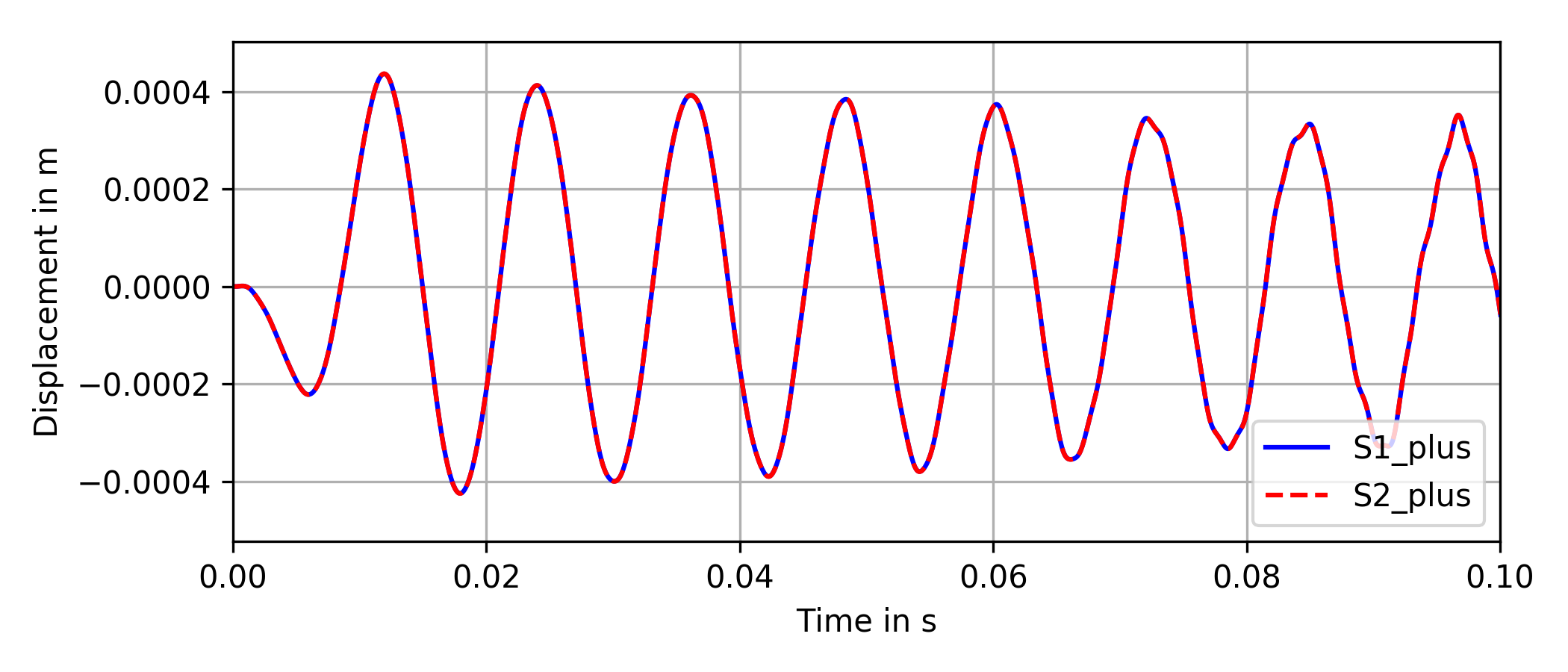}
\caption{Displacement signal of sensor position S1\_plus and  S2\_plus.}
\label{fig:disp_signal}
\end{figure}

\begin{figure}[htb]
\centering
\includegraphics[width=0.8\textwidth]{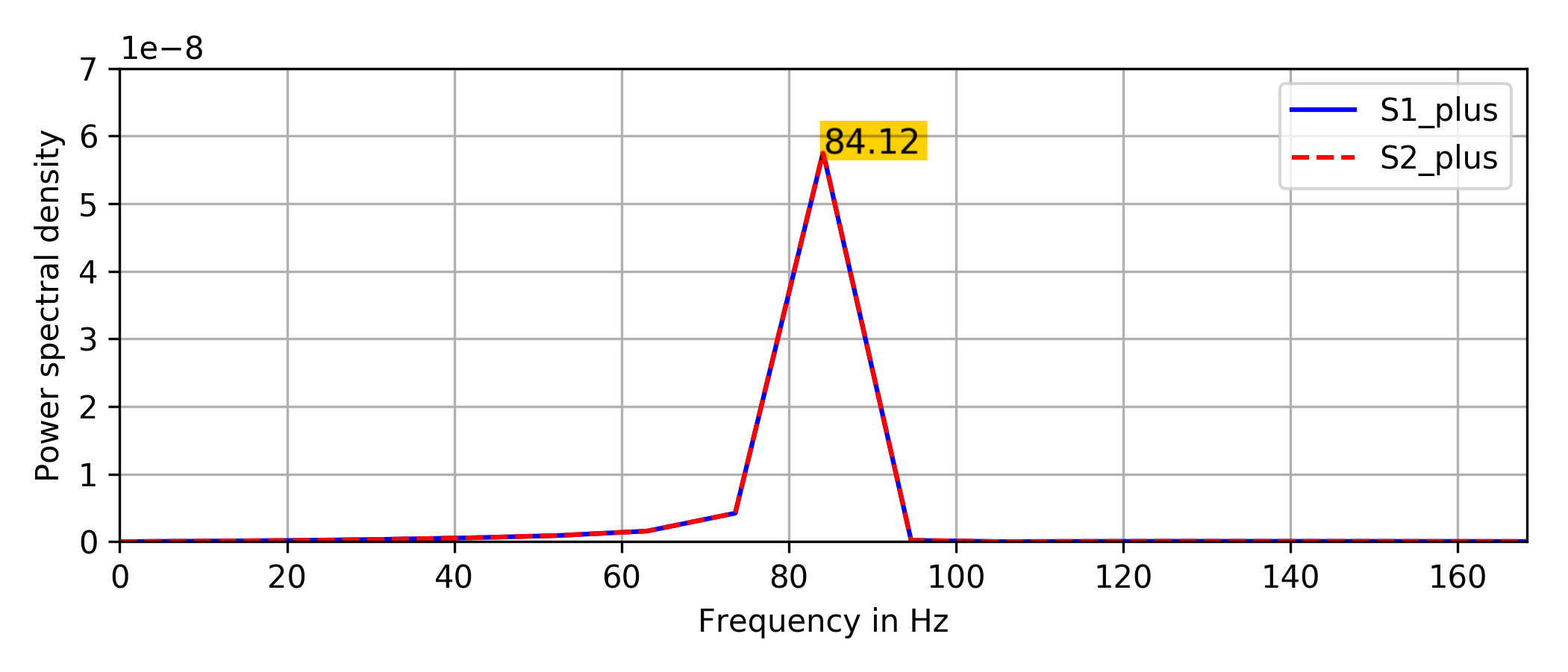}\\
\caption{Discrete Fourier analysis of the displacement signal.}
\label{fig:fft_disp}
\end{figure}

\begin{figure}[htb]
\centering
\includegraphics[width=0.8\textwidth]{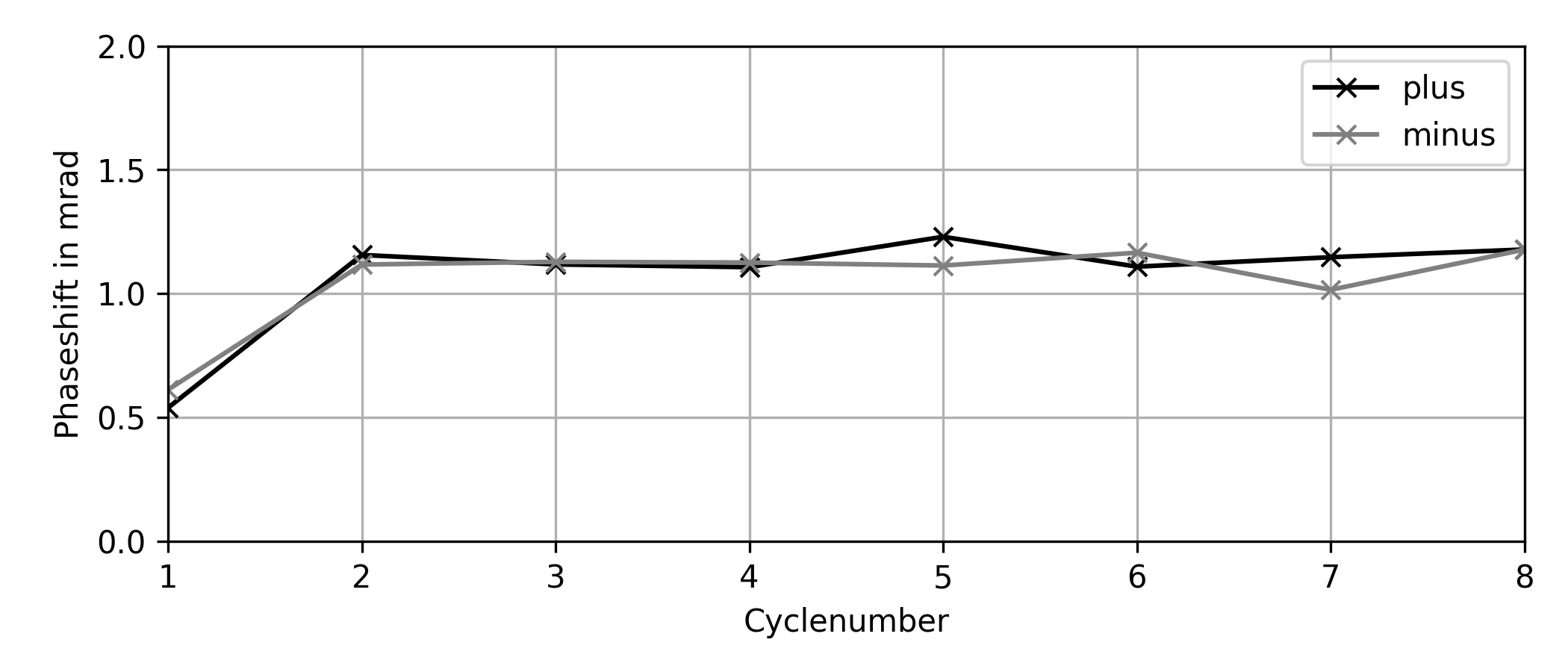}\\
\caption{Phaseshift of the displacement signal.}
\label{fig:phase_disp}
\end{figure}

After the modal analysis has determined the Eigenfrequency of the pipes filled with water, the transient fluid structure simulation is used to extract the phase shift.
Firstly, the fluid field is initialized according to Section~\ref{sec:fluid_boundary}. 
The simulation procedure, which is described in Section~\ref{sec:sim_proced}, is executed in every coupling step $\Delta t_c$. 
The coupling period is chosen to the fluid time step $\Delta t^f$ to minimize the time shift problem of the staggered approach. 
The simulation takes a total of 15 cycles which corresponds to approx. $0.18\,\mathrm{s}$ at the Eigenfrequency. 
Every cycle consists of 202 coupling steps. The displacement signal is extracted at the sensor positions S1\_plus, S1\_minus, S2\_plus and S2\_minus, where plus and minus indicate the left and right measuring pipe, respectively.
The written data files are post processed to extract the phase shift and the frequency of the displacement signals.
The displacement signals of sensor S1\_plus and  S2\_plus are depicted in Figure~\ref{fig:disp_signal}.
It can be seen that the signal is almost sinusoidal in the first 5 cycles and the amplitude slowly decays over time.
The last depicted periods show irregularities and differ from the expected pure sinusoidal course of the displacement signal.
Furthermore a frequency analysis is performed to estimate resonance frequency. 
The results can be seen in Figure~\ref{fig:fft_disp}. 
The highest peak at $84.12\,\mathrm{Hz}$ in the frequency analysis is in good agreement with the estimated excitation frequency.
A discrete Hilbert transformation is applied on the displacement signal to calculate the phase shift, see Figure~\ref{fig:phase_disp}. 
The stability of the FSI system is given for the first 8 cycles of the simulation. 
The symmetry condition, which should be fulfilled due to a axial-symmetric geometry, is only slightly violated.
The error of the averaged phase shift value $\phi_{sim}$ with respect to the experimental data $\phi_{ref}$ is smaller than 5\%, which is shown in Table~\ref{tab:phaseshift_error}. 
\begin{table}[htb]
\centering
  \caption{Phaseshift errors for different mass flows and coupling steps per period.} 
  \begin{tabular}{@{}ccccccccc@{}}
    \toprule
    Mass flow in $\mathrm{\frac{kg}{h}}$ & $\phi_{sim}$ in $\mathrm{mrad}$ & $\phi_{ref}$ in $\mathrm{mrad}$ & Error in $\%$ & Coupling steps\\
    \midrule
$\numprint{20000}$ & - &	0.62 & Instable & \phantom{0}51\\
$\numprint{20000}$ & - &	0.62 & Instable & 101\\
$\numprint{20000}$ & 0.59 &	0.62 & 4.7 & 202\\
$\numprint{40000}$ & 1.18 &	1.23 & 4.1 & 202\\
    \bottomrule
  \end{tabular}
  \label{tab:phaseshift_error}
\end{table}
The relative errors for a mass flow of $\numprint{20000}$ and $\numprint{40000}\,\mathrm{\frac{kg}{h}}$ are less than 5\%. 
Numerical experiments with a lower amount of coupling steps are diverging in the first period, which indicates that the reduction of coupling steps does not lead to satisfactory results. Consequentially, 202 coupling periods are necessary to stabilize the simulation.\\
The simulation runtime was evaluated on a single node which consists of two deca-core Intel Xeon E5-2660 v3 processors. 
The comparison of the runtime to other numerical FSI simulations is depicted in Table~\ref{tab:compare_tfsi}. 
\begin{table}[htb]
\centering
\caption{Comparison of computation time between the present approach to literature values.} 
  \begin{tabular}{@{}lcccccccc@{}}
    \toprule
    Study & Coupling resolution & Periods & Computation time in h\\
    \midrule
Bobovnik et al. (2013)~\cite{Bobovnik.2013}  & \phantom{0}70 & 43 & 72-96 &\\
Kumar et al. (2011)~\cite{Kumar.2011}  & \phantom{0}20 & 15 & 60 &\\
Mole et al. (2008)~\cite{Mole.2008}  & 140 & 15 & 72 &\\
Present & 202 & 15 & 65 &\\
    \bottomrule
  \end{tabular}
\label{tab:compare_tfsi}
\end{table}
It can be seen that both computation time and calculated periods of the present study are comparable to literature values. 
The computation runtime is estimated to 65 hours and over 3000 coupling steps are performed. 
It is noticeable that the computing time has hardly changed over the years. 
This is a consequence of the segregated approach, if two solvers are involved in the FSI approach.
The partitioning of the fluid and the solid domain differs due to the numerical method and geometrical constraints. 
This implies that the exchanged information are collected and communicated between the solvers, which is a time consuming step that is very difficult to parallelize.

\section{Conclusion and Outlook}
\label{sec:conclusion}
An FSI approach was presented for the simulation of a CMF.
Thereby, the open source framework OpenLB and Elmer were used to create a segregated approach. The target equations of the structural and fluid domain were described. In addition, the coupling conditions and the implementation were outlined in detail.
The FEM mesh generation process utilized the open source meshing tool Gmsh to ensure a complete open source workflow. 
A modal analysis was performed to extract the excitation frequency of water and air conveying pipes.
The found excitation frequency was in good agreement to experimental measurements (error $\approx3\%$).
Afterwards, the FSI simulation, which uses the determined excitation frequency, was executed.
The FSI simulation was stable for several cycles and allows to extract the phase shift with a sufficient precision (error $\approx 5\%$).
Therefore, the presented FSI approach for CMF is able to describe the operating principle of a CMF.
Furthermore, the runtime time of the created FSI coupling were comparable to literature approaches using commercial software. 

Nevertheless, certain issues should be addressed in future studies. 
The FSI simulation becomes unstable after several periods.
The reasons for this upcoming instability could be diverse.
Firstly, the coupling time step could be decreased to reduce the time shift problem of the staggered coupling approach.
Unfortunately this leads to an extended calculation time. 
Another possibility is the introduction of a subiteration scheme~\cite{heil2004} that reduces the added mass effect due to the time shift. 
Further improvements can be made by the calculation of the hydrodynamic force, because momentum exchange based approaches suffer from inaccuracy, if too few points are used for integration.
Therefore, a stress based calculation proposed in Geller et al.~\cite{geller2006} may be an alternative.
Furthermore, the applied linear mapping method between the uniform Cartesian LBM grid and the unstructured FEM grid can be improved by using more complex mapping methods~\cite{geller2011}.\\

\bibliographystyle{unsrtnat}
\bibliography{references}

\end{document}